\documentclass[
 reprint,
 amsmath,amssymb,
 aps,
 prd,
 nofootinbib,
 superscriptaddress,
]{revtex4-2}

\usepackage{multirow}
\usepackage{adjustbox}
\usepackage{orcidlink} 
\usepackage{graphicx}   
\usepackage{dcolumn}    
\usepackage{bm}         
\usepackage{hyperref}   
\usepackage{xcolor}
\usepackage[normalem]{ulem}

\hypersetup{
    colorlinks  = true,
    linkcolor   = teal,    
    citecolor   = teal,    
    urlcolor    = teal,    
    filecolor   = teal,    
    anchorcolor = teal    
}

\newcommand{\Msol}{\,\text{M}_{\odot}}          
\newcommand{\MeV}{\,\text{MeV}}                 
\newcommand{\MeVfm}{\MeV\,{\rm fm}^{-2}}        
\newcommand{\fm}{\,{\rm fm}}                    

\newcommand{\nsat}{n_{\rm sat}}                 
\newcommand{\nBH}{n_B^H}                        
\newcommand{\nBc}{n_{B,c}}                      
\newcommand{\nBcH}{n_{B,c}^H}                   
\newcommand{\YLH}{Y_L^H}                        
\newcommand{\YiH}{Y_i^H}                        
\newcommand{\SH}{S_H}                           
\newcommand{\PH}{P_H}                           

\newcommand{\Tcen}{T_{\rm c}}                 
\newcommand{\TcCFL}{T_c^{\rm CFL}}              
\newcommand{\Tnuc}{T_{\rm nuc}}                 
\newcommand{\TQs}{T_{\Qs}}                      

\newcommand{\Qunp}{Q_{\rm unp}}                 
\newcommand{\Qcfl}{Q_{\rm CFL}}                 
\newcommand{\Qs}{Q*}                            
\newcommand{\Rs}{R_*}                           
\newcommand{\Ws}{W_*}                           
\newcommand{\Vs}{V_*}                           
\newcommand{\As}{A_*}                           
\newcommand{\Rsunp}{R_*^{\rm unp}}              
\newcommand{\Rscfl}{R_*^{\rm CFL}}              
\newcommand{\Wsunp}{W_*^{\rm unp}}              
\newcommand{\Wscfl}{W_*^{\rm CFL}}              
\newcommand{\Wunp}{W_{\rm unp}}                 
\newcommand{\Wcfl}{W_{\rm CFL}}                 
\newcommand{\Wbulk}{W_*^{\rm bulk}}             
\newcommand{\Wfin}{W_*^{\rm finite\,size}}      
\newcommand{\Wcoul}{W_*^{\rm Coulomb}}          
\newcommand{\dnC}{\delta n_C}                   

\newcommand{\Dzero}{\Delta_0}                   
\newcommand{\Rgap}{R_\Delta}                    

\newcommand{\Gzero}{\Gamma_0}                   
\newcommand{\Omegazero}{\Omega_0}               
\newcommand{\taudyn}{\tau_{\rm dyn}}            
\newcommand{\sigcrit}{\sigma_{\rm crit}}        

\newcommand{\Bq}{B^{1/4}}                       
\newcommand{\alphas}{\alpha_s}                  
\newcommand{\alphaem}{\alpha_{\rm em}}          
\newcommand{\ms}{m_s}                           

\newcommand{\MB}{M_B}                           
\newcommand{\Mmax}{M_{\rm max}}                 
\newcommand{\MBmax}{M_{B,\max}}                 
\newcommand{\MmaxNS}{M_{\rm max}^{\rm NS}}      
\newcommand{\MmaxQS}{M_{\rm max}^{\rm QS}}      
\newcommand{\MmaxPNS}{M_{\rm max}^{\rm PNS}}    
\newcommand{\MPNS}{M^{\rm PNS}}             
\newcommand{\MQS}{M^{\rm QS}} 
\newcommand{\Econv}{E_{\rm conv}}               

\newcommand{\tzero}{t_0}                        
\newcommand{\tTmax}{t_{T_{\max}}}               

\begin{document}

\preprint{}

\title{Coexistence of strange quark stars and neutron stars:\\metastability and nucleation in proto-neutron stars}

\author{Mirco Guerrini
\orcidlink{0009-0009-5946-3900}}
\email{mirco.guerrini@unife.it}
\affiliation{Department of Physics and Earth Science, University of Ferrara, 44122 Ferrara, Italy}
\affiliation{INFN, Sezione di Ferrara, 44122 Ferrara, Italy}

\author{Giuseppe Pagliara
\orcidlink{0000-0003-3250-1398}}%
\affiliation{Department of Physics and Earth Science, University of Ferrara, 44122 Ferrara, Italy}
\affiliation{INFN, Sezione di Ferrara, 44122 Ferrara, Italy}

\author{Luca Passarella
\orcidlink{0009-0005-0132-7579}}%
\affiliation{Department of Applied Science and Technology, Politecnico di Torino, 10129 Torino, Italy}
\affiliation{INFN Sezione di Torino, 10125 Torino, Italy}

\author{Alessandro Drago
\orcidlink{0000-0003-1302-8566}}%
\affiliation{Department of Physics and Earth Science, University of Ferrara, 44122 Ferrara, Italy}
\affiliation{INFN, Sezione di Ferrara, 44122 Ferrara, Italy}


\begin{abstract}
If strange quark matter (SQM) is absolutely stable, hadronic neutron stars (NSs) and strange quark stars (QSs) may coexist in the so-called two-families scenario. A key issue is explaining how hadronic NSs can survive as long-lived metastable objects, rather than promptly converting into stable QSs. Since all NSs are born as hot proto-neutron stars (PNSs), a necessary condition for their existence is that the conditions for conversion, triggered by the nucleation of the first critical SQM droplet, are not reached in every PNS.
We investigate the thermal nucleation of critical SQM droplets under representative PNS conditions, accounting for flavor-composition fluctuations and finite-size effects on color superconductivity by suppressing pairing in droplets smaller than the diquark coherence length. We define the nucleation conditions as the thermodynamic state at which a critical SQM droplet is expected to nucleate statistically within the characteristic dynamical timescale of the system.
Under these conditions, the reduced nucleation barrier is nearly universal, with $\Ws/T\simeq167$ to within a few units. Exploring the SQM parameter space of a bag-like model, with a fixed hadronic EOS compatible with a recent Bayesian analysis of the two-families scenario, we find parametrizations and surface tensions for which a canonical $1.4\Msol$ NS survives the PNS stage, a necessary condition for the two-families scenario. For this specific combination of hadronic and quark models, requiring such survival sets a lower bound on the hadron--quark surface tension, $\sigma\gtrsim 67\MeVfm$.
\end{abstract}

\maketitle

\section{Introduction}
\label{sec:intro}

The Bodmer--Witten hypothesis (BW)  posits that bulk three-flavor strange quark matter (SQM), composed of up, down, and strange quarks in weak equilibrium, is absolutely stable: at zero pressure and temperature its energy per baryon would lie below that of the most stable nucleus, $\left.(E/N_B)\right|_{\rm SQM} < \left.(E/N_B)\right|_{^{56}\rm Fe} \simeq 930\MeV$~\cite{Bodmer:1971we,Witten:1984rs}. The consequences of such a hypothesis are far-reaching and would provide signatures in astrophysics and cosmology~\cite{Alcock:1986hz,Madsen:1998uh} (see~\cite{DiClemente:2024lzi} for a recent reappraisal).

The BW hypothesis does not conflict with the existence of ordinary nuclei. Converting a nucleus into SQM would require the near-simultaneous weak conversion of a large number of light quarks into strange quarks, an extremely high-order process with negligible rate. Moreover, small strangelets are energetically penalized by finite-size effects, so that there exists a minimum baryon number below which SQM lumps are unstable~\cite{Madsen:1998uh,Weber:2004kj}. The path in configuration space is thus not downhill, and ordinary hadronic matter is \emph{metastable}: it sits in a local minimum of the free energy, is protected from the true SQM ground state by an energy barrier, and has a lifetime far exceeding the age of the Universe.

The same reasoning, applied to compact stars, is far less restrictive. If SQM is absolutely stable, every neutron star (NS) is in principle metastable against conversion to a strange quark star (QS)~\cite{Alcock:1986hz}. Two outcomes are then possible: either the conversion conditions are generically met during stellar evolution, so that \emph{all} compact stars are QSs, or they are met only in part of the population, so that NSs and QSs coexist as assumed within the so-called two-families scenario~\cite{Drago:2013fsa,Drago:2015cea}. The all-QS option is disfavored because some observed phenomenology, such as glitches~\cite{Pizzochero:2019hby,Haskell:2018rvj,Basu:2018fbt,Poli:2023vyp} and quasi-periodic oscillations~\cite{Watts:2006hk}, is naturally accounted for by the nuclear crust of NSs, and not by QSs, which are either bare or endowed with a crystalline color-superconducting crust whose properties differ from those of an ordinary nuclear crust~\cite{Anglani:2013gfu,Mannarelli:2014ija,Mannarelli:2015jia}. It is therefore more plausible that at least a fraction of compact stars are hadronic objects, as realized in the two-families scenario. 
In this respect, the two-families scenario shares some similarities with a well-known fact in nuclear astrophysics: not all main-sequence stars are able to synthesize iron (the most stable nucleus) in their cores, but only those which are massive enough. Stars with different internal compositions therefore coexist. Also, the idea of metastability is not at all new in nuclear astrophysics: most of the Universe is made of hydrogen and only under specific conditions (which are met inside stars) hydrogen can undergo fusion reactions leading to the formation of more stable nuclei.

The scenario of coexistence is further motivated by the tension between the stiffness requirements that different observations impose on the equation of state (EOS). On the one hand, the existence of very massive compact stars~\cite{Romani:2022jhd,Antoniadis:2013pzd,Fonseca:2021wxt,Riley:2021pdl,LIGOScientific:2020zkf} calls for a stiff EOS. On the other hand, the appearance of hyperons and $\Delta$ isobars softens the EOS~\cite{Vidana:2000ew}, and several observations, such as heavy-ion flow data~\cite{Danielewicz:2002pu}, the tidal deformability inferred from GW170817~\cite{LIGOScientific:2018cki}, and the small radii measured at ordinary compact-star masses ($\sim 1.4\Msol$)~\cite{Mauviard:2025dmd,Ozel:2016oaf}, point instead to a soft EOS. In the two-families scenario this tension is resolved by decoupling the two regimes: the soft hadronic branch accounts for the canonical-mass small-radius objects, while the stiff SQM branch accounts for the most massive ones. This picture has recently received quantitative support from the first Bayesian analysis of the two-families scenario against astrophysical data~\cite{Passarella:2026mxy}. 
The scenario also offers an interpretation of subsolar-mass compact objects as QSs\footnote{We stress that the difficulty in interpreting HESS~J1731--347, and subsolar-mass compact objects in general, as NSs is not related only to their mass and radius, but also to the lack of a robust formation channel for NSs with masses below $\sim 1.17$--$1.19\Msol$~\cite{Suwa:2018uni,Muller:2024aod}. A possible formation mechanism for subsolar-mass QSs has instead been proposed in~\cite{DiClemente:2022wqp}.}, such as the compact object associated with HESS~J1731--347~\cite{Doroshenko:2022nwp,DiClemente:2022wqp} or the sources of the candidate subsolar-mass gravitational-wave events~\cite{LIGOScientific:2018glc,LIGOScientific:2021job,Morras:2023jvb}; see~\cite{Gao:2026zwc} for the first numerical simulation of subsolar-mass QS--QS mergers.  

A central open question for the two-families scenario then concerns the metastability itself: can NSs persist as long-lived metastable objects without decaying into QSs, and how is the QS branch actually populated? Answering this question requires identifying the thermodynamic conditions under which a hadronic object converts into a QS, and the astrophysical systems in which those conditions are naturally realized. Specifically, the very existence of the two families demands that during their lifetime at least some compact stars survive as long-lived metastable NSs while others convert into QSs.

Assuming that deconfinement at high density is a first-order phase transition, a nonzero hadron--quark surface tension $\sigma$ separates the two phases and gives rise to a free-energy barrier. As a consequence, the system can remain trapped in the metastable local minimum even when a more stable phase exists, instead of decaying into it instantaneously. The decay occurs only when a spontaneous thermal or quantum fluctuation creates a droplet of the stable phase which is large enough to overcome the free-energy barrier, i.e. large enough to keep expanding: this is the process of \emph{nucleation}\footnote{Another possible mechanism for a first-order phase transition is spinodal decomposition, which occurs when the system enters a mechanically unstable region before nucleation triggers the phase transition because of a quench in the thermodynamical conditions~\cite{Bessa:2008nw}. A further discussion is provided in Sec.~\ref{sec:conclusions}.}~\cite{Olesen:1993ek,Langer:1969bc,Iida:1997ay}. Once a critical droplet is formed, the conversion proceeds through subsequent stages: the droplet grows, individual bubbles grow and coalesce\footnote{It has recently been argued that the coalescence of nucleated bubbles following a first-order transition, for instance in binary mergers or core-collapse supernovae, can source gravitational waves (GWs) in the MHz band~\cite{Blas:2022xco,Bleau:2026ala,Ecker:2026ixk}, providing an additional motivation for a quantitative treatment of the nucleation process.}, until reaching a macroscopic size, after which a combustion front is formed, eventually converting the entire star~\cite{Olinto:1986je,Pagliara:2013tza,Herzog:2011sn,Drago:2015fpa,Furusawa:2015gjj,Furusawa:2015vfn}.

In many studies of the two-families scenario, the nucleation conditions are identified through a simple geometric criterion~\cite{DePietri:2019khb,Becerra:2025ryy,Passarella:2026mxy}: the conversion is assumed to be triggered once the mean distance between strange quarks in the hadronic phase becomes comparable to the average inter-nucleon distance in nuclear matter, i.e. once the strange quarks are close enough to mutually interact and seed the first droplet of deconfined matter. While simple and effective, this is a qualitative proxy rather than the outcome of an explicit evaluation of the nucleation rate; in this work we instead adopt a microphysical treatment of the nucleation process.

Previous literature has explored various aspects of nucleation in astrophysics, employing a range of underlying EOSs, environmental conditions (such as zero or finite temperature, and the presence or absence of trapped neutrinos), and assumptions about the composition of the nucleating droplet, which is determined by the interplay between deconfinement and weak reaction timescales~\cite{Iida:1997ay,Iida:1998pi,Olesen:1993ek,Bombaci:2009jt,Bombaci:2011mx,Mintz:2009ay,Mintz:2010mh,Ren:2020tll,Horvath:1992wq,Horvath:1994zn,Fraga:2005wgs,Lugones:2011zz,Logoteta:2012ms,Logoteta:2012zz,Bombaci:2016xuj,Lugones:2015bya,Lugones:2011xv}. Building on this, it was recently examined how thermal fluctuations of the conserved charges (i.e., flavor composition) impact the nucleation of two-flavor quark matter~\cite{Guerrini:2024gzu}. The formalism developed there has been 
extended to the case of three-flavor matter~\cite{Guerrini:2025mxx}.

Another key microphysical ingredient one has to consider in this context is related to the phenomenon of color superconductivity. At asymptotically large baryon density the color-flavor-locked (CFL) phase is the energetically favored ground state of three-flavor quark matter~\cite{Alford:2007xm}; its relevance at the densities reached in compact stars is less certain, but since pairing lowers the energy of quark matter, a color-superconducting phase may be a good candidate for being the absolutely stable SQM within the BW hypothesis. This raises a problem specific to nucleation: the diquark coherence length~\cite{Alford:2007xm} is comparable to, or larger than, the size of the earliest subcritical droplets, so that the smallest seeds cannot sustain a fully developed condensate and are predominantly unpaired. The color-superconducting phase can develop only once the droplets are large enough to allow the formation of diquark pairs~\cite{Alford:2007xm,Amore:2001uf}. While past literature has included color-superconducting EOSs to evaluate nucleation rates~\cite{Carmo:2013fr,Vidana:2008zz,Bombaci:2006cs}, the impact of incomplete diquark-pair formation in small systems has generally been neglected. Here, we account for this phenomenon through a simple scheme in which the thermodynamics of the quark phase depends on the droplet radius.

The aim of this work is to assess whether SQM nucleation conditions are reached during the early evolution of a hadronic proto-neutron star (PNS), producing QSs. Specifically, we compute the nucleation rate under typical PNS conditions and investigate whether, for physically motivated values of the model parameters, the resulting timescales prevent nucleation in a fraction of PNSs, which then
can survive as metastable NSs once cooled. 
Since all NSs are born as PNSs, their existence requires that not all PNSs convert into QSs. However, note that QSs may also form through other channels, such as in binary compact-star mergers (BCSMs)~\cite{DePietri:2019khb}, during a core-collapse supernova (CCSN), or later during accretion or spin-down of a cold NS~\cite{Becerra:2025ryy}. The absence of nucleation in at least some hadronic PNSs is therefore a necessary, but not sufficient, condition for the two-families scenario.

The paper is organized as follows. In Sec.~\ref{sec:eos} we introduce the EOS models used for the hadronic and quark phases. In Sec.~\ref{sec:pns} we describe the approximate thermodynamic evolution of a PNS. In Sec.~\ref{sec:conversion} we present the nucleation formalism, including the finite-size treatment of color superconductivity, and its role in triggering the conversion into a QS. Our results are presented and discussed in Sec.~\ref{sec:results}, and we summarize our conclusions in Sec.~\ref{sec:conclusions}. We use natural units $\hbar=c=k_B=1$.

\section{Equations of state}
\label{sec:eos}

A microscopic QCD description of strongly interacting matter across the relevant range of densities, temperatures, and conserved-charge content is not yet available, as QCD remains intractable at finite baryon chemical potential. We therefore adopt the standard strategy of describing the confined hadronic phase and the deconfined quark phase using two separate effective models. 

It is worth emphasizing that, in scenarios where SQM is not absolutely stable, compact stars with and without a deconfined quark matter  core are governed by the same equilibrium EOS. Under the BW hypothesis, however, the hadronic and the SQM branches are described by two distinct EOSs, corresponding to a metastable and an absolutely stable solution of QCD, respectively. 

In this section, we summarize the characteristics of the EOS models employed in this work; see~\cite{Guerrini:2025ojt} and references therein for a more complete description. 

Electrons, positrons, electron neutrinos, and antineutrinos are treated as non-interacting fermions with masses $m_e=0.511\MeV$ and $m_\nu=0\MeV$, and degeneracies of 2 and 1, respectively. We also include thermal photons, treated as an ideal boson gas with a degeneracy of 2.

\subsection{Hadronic EOS}
\label{sec:eos_had}

The hadronic phase is described by the relativistic mean-field (RMF) model SFHo-H$\Delta$, in which baryons interact through the exchange of the scalar $\sigma$ and vector $\omega,\phi,\rho$ mesons. It is built on the SFHo parametrization~\cite{Steiner:2012rk}, extended to the full baryon octet~\cite{Fortin:2017dsj} and the $\Delta$-isobar quartet~\cite{Drago:2014oja}, and has been widely employed in the two-families scenario~\cite{Drago:2015cea,Bombaci:2020vgw,DePietri:2019khb,Becerra:2025ryy}. 

The nucleonic sector follows~\cite{Steiner:2012rk}; the hyperon couplings are fixed by SU(6) relations for the vector mesons~\cite{Fortin:2017dsj} and by fitting the scalar couplings to the empirical potential depths $U_\Lambda=-28\MeV$, $U_\Sigma=+30\MeV$, and $U_\Xi=-18\MeV$~\cite{Schaffner:1993nn,Schaffner:1995th}. The $\Delta$ couplings are fixed with $x_{\sigma\Delta}=g_{\sigma\Delta}/g_{\sigma N}=1.15$ and $x_{\omega\Delta}=x_{\rho\Delta}=1.0$. Light pseudoscalar mesons are added as a thermal ideal Bose gas with effective chemical potentials tied to the same mean fields~\cite{Moller:1996uf,Lavagno:2010ah}.

This hadronic EOS is compatible with the two-families posterior of the recent Bayesian analysis of~\cite{Passarella:2026mxy}, which constrains a closely related nonlinear RMF model with hyperons and $\Delta$ resonances against astrophysical observations, heavy-ion collision data and chiral effective field theory calculations.
In this work the hadronic EOS is kept fixed; the expected impact of a different hadronic parametrization is discussed qualitatively in Sec.~\ref{sec:results_paramspace}.

The EOS model presented here is employed for the nuclear and supra-nuclear regimes in NSs and hadronic PNSs. For the sub-nuclear regime, corresponding to the NS crust and the PNS envelope, we adopt the \textsc{Compose} table\footnote{\url{https://compose.obspm.fr/eos/118}}~\cite{Fortin:2017dsj,Steiner:2012rk,Moller:1996uf,Hempel:2009mc}. 

We use $(\nBH,\{\YiH\},T)$ as free variables, where $\nBH$ is the baryon density of the hadronic phase, $\YiH=n_i^H/\nBH$ is the net fraction of particle $i$ (with $i=p,n,\Lambda,\Sigma^{+,0,-},\Xi^{0,-},\Delta^{++,+,0,-},e,\nu_e$), and $T$ is the temperature. 
Equivalently, the composition can be labelled by the net fractions of the charges conserved by the strong interaction, $Y_B^H\equiv1$, $Y_C^H$ and $Y_S^H$, i.e.\ the net electric charge (excluding leptons) and net strangeness per baryon of the hadronic phase.
Note that under the assumptions of chemical equilibrium and charge neutrality, all fractions $\{\YiH\}$ can be written as a function of $\nBH, \YLH, T$ (where $\YLH=Y_e^H+Y_{\nu_e}^H$ is the lepton fraction) in the case in which neutrinos are trapped, or simply as a function of $\nBH, T$ if neutrinos are untrapped and therefore their net number is vanishing: ($Y_{\nu_e}^H=0$).

\subsection{Quark EOS}
\label{sec:eos_quark}

We consider two deconfined SQM phases: the unpaired phase when the system is smaller than the typical diquark pair coherence length, and the CFL phase when it is larger (see Sec.~\ref{sec:conversion_nucleation_unpcfl}). The deconfined phase is described by the $\alpha$Bag model, an MIT bag model supplemented by leading-order perturbative QCD corrections in the strong coupling $\alphas$, which is used as a free parameter~\cite{Weissenborn:2011qu,Farhi:1984qu}. The finite-temperature generalization follows the prescription of~\cite{Fischer:2010wp}, combining the kinetic contributions of each massive quark flavor with the analytic one-gluon exchange corrections of massless quarks; see~\cite{Raza:2026imv} for a recently developed alternative.

In the unpaired case ($\Qunp$), the thermodynamic quantities are obtained by summing the single-flavor contributions and shifting the pressure and energy density by the bag constant $B$, which encodes the difference in energy density between the perturbative and non-perturbative QCD vacua. In the CFL phase ($\Qcfl$), the pairing contribution is added on top of the unpaired EOS through a term depending on the pairing gap $\Delta(T)$, similarly to~\cite{Alford:2002rj,Weissenborn:2011qu}.

The temperature dependence of the gap is modeled as~\cite{Schmitt:2010pn,Alford:2007xm}
\begin{equation}
    \Delta(T) = \Theta(\TcCFL-T)\,\Dzero\,\sqrt{1-\left(\frac{T}{\TcCFL}\right)^2},
    \label{eq:gap}
\end{equation}
where $\TcCFL$ is the critical temperature
\begin{equation}
    \TcCFL \simeq 2^{1/3}\times 0.57\,\Dzero.
\end{equation}
This ansatz reproduces the qualitative features of the gap obtained from the gap equation in the weak-coupling limit: a zero-temperature value $\Dzero$, a mean-field square-root vanishing near the critical temperature, and a BCS-like relation between $\TcCFL$ and $\Dzero$~\cite{Alford:2007xm,Schmitt:2010pn}. 

In the unpaired phase the composition is not fixed a priori: the flavor fractions $Y_i^{\Qunp}$ are independent degrees of freedom that can be determined by imposing suitable physical conditions. In the CFL phase, by contrast, the requirement of equal quark Fermi momenta enforces $n_u^{\Qcfl}=n_d^{\Qcfl}=n_s^{\Qcfl}$, which locks the composition to $Y_C^{\Qcfl}=0$ and $Y_S^{\Qcfl}=1$; the flavor fractions are therefore no longer free.

The current quark masses are fixed to $m_u=m_d=0$ and $\ms=100\MeV$. We treat $B$, $\Dzero$, and $\alphas$ as free parameters, specifying their values explicitly when presenting our results. See~\cite{Guerrini:2025ojt} for the explicit expressions of the thermodynamic quantities.

\section{Evolution of proto-neutron stars}
\label{sec:pns}

Hadronic PNSs are one of the possible remnants of CCSNe~\cite{Prakash:1996xs}.
When stars with initial masses above $\sim 8\Msol$ exhaust their nuclear fuel, their cores can no longer be supported against gravity and begin to collapse.
The collapse proceeds until the core baryon density approaches nuclear saturation, $n_B\simeq \nsat \simeq 0.16~\mathrm{fm}^{-3}$, where the repulsive component of the nuclear force together with neutron degeneracy pressure halts the collapse, causing the core to bounce. 

Immediately after bounce, the hadronic PNS is formed.
It is characterized by a relatively low mass $\sim(0.5$--$1)\Msol$, a high lepton fraction $Y_{L}\simeq 0.35$--$0.4$, and an entropy per baryon of $S/N_B\simeq 1$--$1.5$ (with precise values depending on the progenitor, on the EOS, and on weak-interaction microphysics)~\cite{Prakash:1996xs,Fischer:2010wp,Fischer:2017lag,Kuroda:2021eiv}.
Over the subsequent $\sim(0.5$--$1)$~s, the PNS accretes matter from the overlying layers. The amount of accreted mass depends on the progenitor star characteristics (mainly its mass). 

If accretion is modest (e.g., for progenitors $\lesssim20\Msol$), a neutrino-driven CCSN explosion happens, leaving a hadronic PNS remnant.
For more massive progenitors, sustained accretion can be large enough that conditions for quark deconfinement may already be met during this early phase. Indeed, deconfinement has been proposed as a mechanism that can trigger CCSN explosions in cases where the standard mechanism fails~\cite{Fischer:2017lag,Kuroda:2021eiv}. This possibility, in the context of the two-families scenario, will be investigated in a forthcoming paper. 
For even more massive progenitors, continued accretion can drive the PNS above its stability limit, leading to black hole (BH) formation.
In the present work we focus on the scenario in which a hadronic PNS survives these initial stages and becomes the remnant of a successful CCSN.

After explosion, the newly formed hadronic PNS core (typically $S/N_B\simeq 1$--$1.5$ and $Y_{L}\simeq 0.35$--$0.4$) is no longer strongly accreted, so its baryonic mass $\MB$ can be taken as approximately constant for the subsequent evolution.
Within $\sim10$ s, the core entropy per baryon commonly rises toward $S/N_B\simeq 1.5$--$2$, while the lepton fraction decreases to $Y_{L}\simeq 0.25$--$0.3$ owing to neutrino diffusion (deleptonization).
Finally, the PNS cools by neutrino emission, eventually reaching the cold neutrinoless $\beta$-equilibrium NS configuration after a few tens of seconds.

The specific values of $S/N_B$ and $Y_{L}$ depend on the PNS baryonic mass (hence on progenitor properties), on the EOS and on the neutrino-transport treatment adopted~\cite{Pons:1998mm,Pons:2001ar,Roberts:2012um,Roberts:2012zza,Camelio:2017nka,Camelio:2017hrs}.

\section{Conversion into a strange quark star}
\label{sec:conversion}

The conversion of a hadronic PNS into a QS can be schematically divided into several steps~\cite{Drago:2015dea}:  
(i) \emph{Nucleation}: a droplet of SQM large enough to keep expanding is formed;  
(ii) the first seed of SQM expands further or merges with other droplets until reaching a macroscopic size;  
(iii) the subsequent propagation of the conversion front can be described hydrodynamically and proceeds in two phases: an initial rapid burning (lasting a few milliseconds), whose velocity is greatly enhanced by hydrodynamical instabilities, which converts the central region of the star~\cite{Pagliara:2013tza,Furusawa:2015gjj,Drago:2015fpa}; then a slower burning regime (lasting tens of seconds), driven by the production and diffusion of strangeness, converts the outer stellar layers~\cite{Olinto:1986je,Furusawa:2015vfn,Drago:2015fpa}.  

In this work, we focus on identifying the thermodynamic conditions required for conversion to begin, namely under which conditions nucleation occurs, and on comparing them with the typical thermodynamic conditions of a PNS.

\subsection{Nucleation}
\label{sec:conversion_nucleation}

If a phase transition is of first order, a surface tension exists between the two phases. From a thermodynamic point of view, the surface tension is the extra energy per unit area required to create an interface between the two phases, representing the cost of forming a finite-size volume of the new phase rather than an infinite bulk configuration, since particles located at the boundary of a finite volume interact differently from those deep in the interior.

In general, nucleation theory aims at computing the rate at which a critical volume of a new phase $Q$\footnote{In this section, $H$ and $Q$ denote two generic phases of a system; in the physical context considered here, they correspond to the hadronic and deconfined quark phases, respectively.} forms inside a system initially in a metastable phase $H$, thereby triggering the decay of the metastable state through a first-order phase transition.

Let us consider a system of volume $V$ initially in a stable $H$ phase. By increasing density (e.g. raising pressure) or temperature, a different phase $Q$ can become energetically favored, such that the global minimum corresponds to the whole volume being occupied by $Q$\footnote{For simplicity, we neglect here the possibility of a mixed phase being the minimum-energy configuration. This case can be recovered by replacing $V$ with the equilibrium $Q$-phase volume $V_{Q,\mathrm{eq}}\leq V$ in the following discussion. For example, if a pasta phase is the minimum-energy configuration at given conditions, $V_{Q,\mathrm{eq}}$ is the volume of the pasta structures and $V$ the volume of the Wigner--Seitz cell, see e.g.~\cite{Ju:2021hoy}.}. Since the system is initially in a pure $H$ state (a local energy minimum), a finite $Q$ volume $V_Q$ must be created in order to reach the global minimum. However, although the fully converted configuration ($V_Q=V$) is more stable than the pure $H$ phase ($V_Q=0$), intermediate states with $0<V_Q<V$ are not necessarily favored and must still be crossed. In particular, a configuration with volume $V_Q$ is more stable than a pure $H$ state only if $V_Q$ is sufficiently large for the energetically favorable bulk term to outweigh the unfavorable surface contribution.

As a result, there is an energy barrier between the metastable $H$ configuration and the global $Q$ minimum. The maximum of the energy as a function of $V_Q$ defines a \emph{critical volume} $\Vs$ and an \emph{energy barrier height} $\Ws$, the latter being the energy difference between this maximum configuration and the initial local minimum $H$. Namely, $\Vs$ is the critical droplet volume of phase $Q$ in $H$ such that a slightly larger configuration is energetically favored over $V_Q=\Vs$: droplets with $V_Q<\Vs$ tend to shrink, whereas those with $V_Q\geq \Vs$ grow. Thus, a spontaneous local fluctuation producing a droplet at least as large as the critical size ($V_Q\geq \Vs$) is required to overcome the barrier; such a droplet will then grow and eventually lead to the global minimum configuration ($V_Q=V$). These fluctuations can be either thermal or quantum, but we focus here on the thermal case, as it dominates at the typical temperatures of PNSs~\cite{Bombaci:2016xuj,Guerrini:2024gzu}. The spontaneous local fluctuation that generates a critical droplet of the new stable phase is referred to as nucleation.

Finally, it is worth stressing two points. First, the formalism adopted here addresses only the formation of the first critical droplet, i.e. stage (i), and not the subsequent macroscopic evolution triggered by it, i.e. stages (ii) and (iii). Second, we do not address the detailed microphysical mechanism of quark deconfinement responsible for the formation of the deconfined droplet within the hadronic medium; we simply assume that deconfinement occurs on strong-interaction timescales, generating a volume $V_Q$ of deconfined quarks as a spontaneous fluctuation of a thermodynamic variable, and we investigate under which conditions such a fluctuation can be large enough to trigger the macroscopic conversion (see, e.g.,~\cite{Fukushima:2020cmk} for tentative discussions of the microscopic mechanism).

\subsubsection{Fluctuations}
\label{sec:conversion_nucleation_fluctuations}

Besides $V_Q$, the other thermodynamic variables can also fluctuate spontaneously~\cite{Landau:1980mil}. It is therefore crucial to identify which configurations of the system can be explored. In the context of quark nucleation, the spontaneous fluctuations that generate quark droplets through deconfinement are mediated by the strong interaction and occur on the typical timescale $\tau_{\rm strong}\sim 10^{-23}$~s. Because the weak interaction acts on much longer timescales, it cannot play a role during nucleation. As a consequence, only configurations with the same flavor content as the initial state, or equivalently the same baryon number $N_B=\tfrac13 N_u+\tfrac13 N_d+\tfrac13 N_s$, non-leptonic electric charge $N_C=\tfrac23 N_u-\tfrac13 N_d-\tfrac13 N_s$, and strangeness $N_S=N_s$, can be explored during nucleation.

While this consideration is generally accepted in the literature, it is not fully clear whether flavor conservation should be enforced locally or globally (see~\cite{Guerrini:2025mxx,Guerrini:2025ojt} for a discussion). In several works (see e.g.~\cite{Olesen:1993ek,Lugones:2015bya,Iida:1997ay}), the critical droplet of deconfined quark matter is assumed to be in an out-of-chemical-equilibrium state with the same flavor composition as the bulk hadronic phase ($Y_i^Q=\YiH$ for $i=u,d,s$, or equivalently $i=B,C,S$). At finite temperature, however, the flavor composition fluctuates around its bulk average. Since nucleation is a local process, the first critical droplet could form within a subsystem of the hadronic medium where the local composition differs from the average and is more favorable for nucleation. A complete treatment should take this into account, as suggested in~\cite{Norsen:2002qw,DiToro:2006bkw,Mintz:2009ay}. In this work, we adopt the framework introduced in~\cite{Guerrini:2024gzu,Guerrini:2025mxx,Guerrini:2025ojt}, which explicitly incorporates thermal fluctuations of the composition.

As discussed above, if the initial hadronic phase is metastable, a fluctuation in $V_Q$ can lower the energy of the system provided that $V_Q$ is large enough. Conversely, if the initial system is already in chemical and thermal equilibrium with respect to the interaction governing nucleation, a fluctuation in any other variable always increases the total energy. Therefore, the most favorable critical configuration is a saddle point, where the energy is a maximum along the $V_Q$ axis and a minimum along all other directions~\cite{Langer:1969bc,Langer:1974cpa}.

\subsubsection{Energy barrier}
\label{sec:conversion_nucleation_energybarrier}

The energy barrier $\Ws$ is defined as the energy difference between the saddle-point configuration, a critical droplet of SQM (labeled $Q_*$) surrounded by hadronic matter, and the initial configuration of a pure hadronic phase $H$. Equivalently, it represents the work required to generate the critical droplet in a hadronic system~\cite{Landau:1980mil}. Assuming the critical droplet to be spherical with radius $\Rs$, and that its energy can be written as the sum of a bulk term, a finite-size term, and a Coulomb term, following the explicit calculation in~\cite{Guerrini:2025ojt} the energy barrier reads
\begin{eqnarray}
    \Ws &=& \Wbulk + \Wfin + \Wcoul. \label{eq:WWbWfWc}
\end{eqnarray}
The first term is the bulk contribution, proportional to $\Rs^3$, namely a volume term,
\begin{widetext}
\begin{eqnarray}
        \Wbulk &=& -\Vs\left(P_{\Qs}-\PH\right)
        +\sum_{i} N_i^{\Qs}\left(\mu_i^{\Qs}-\mu_i^H\right)
        + S_{\Qs}(\TQs-T) \label{eq:Wbulk}\\
        &=& -\frac{4}{3}\pi \Rs^3\left[\left(P_{\Qs}-\PH\right)
        -n_B^{\Qs}\sum_{i}Y_i^{\Qs}\left(\mu_i^{\Qs}-\mu_i^H\right)
        -s_{\Qs}\left(\TQs-T\right)\right],\notag
\end{eqnarray}
\end{widetext}
where $i=B,C,S,e,\nu_e$ and any thermodynamic quantity $X$ is evaluated using the EOS models of Sec.~\ref{sec:eos} as
\begin{eqnarray}
X_{H}&\equiv& X_H(n_B^{H},\{Y_i^{H}\},T), \\
X_{\Qs}&\equiv& X_Q(n_B^{\Qs},\{Y_i^{\Qs}\},\TQs),
\end{eqnarray}
with $X$ standing for the chemical potentials $\mu_i$, the pressure $P$ and the entropy density $s$. $\Rs$ is the critical radius, namely the radius of the critical droplet. The bulk term is positive if the hadronic phase is stable and negative if it is metastable. As discussed in~\cite{Guerrini:2025mxx,Guerrini:2025ojt}, Eq.~\eqref{eq:Wbulk} assumes that $\Vs$ is much smaller than the surrounding hadronic region, a safe approximation in compact-star conditions (see Sec.~\ref{sec:conversion_nucleation_pns}).

The second term of Eq.~\eqref{eq:WWbWfWc} contains all finite-size corrections, such as the surface term $\propto\sigma \Rs^2$, with $\sigma$ the surface tension, and the curvature term $\propto\varsigma \Rs$, with $\varsigma$ the curvature tension. These quantities should ideally be computed self-consistently from the EOS, and in principle depend on both temperature and density (see, e.g.,~\cite{Lugones:2021tee,Bessa:2008nw}). Different microscopic approaches have been employed in the literature (see, e.g.,~\cite{Bessa:2008nw,Fraga:2018cvr,Kapusta:2007xjq} for field-based methods, and~\cite{Carmo:2013fr,Lugones:2015bya,Lugones:2018qgu,Grunfeld:2024ihq} for thermodynamic approaches based on the multiple-reflection expansion), yielding highly uncertain and model-dependent surface and curvature tensions, ranging from a few units to hundreds of $\MeVfm$~\cite{Lugones:2021zsg,Lugones:2021tee,Heiselberg:1992dx,Iida:1997ay,Alford:2001zr,Bessa:2008nw,Mintz:2009ay,Grunfeld:2024ihq,Palhares:2010be,Pinto:2012aq,Lugones:2015bya,Lugones:2018qgu,Fraga:2018cvr,Schmitt:2020tac,Ju:2021hoy}. Given these uncertainties, we neglect the curvature term and incorporate all finite-size contributions into an effective surface term $\sigma$, treated as a free parameter,
\begin{eqnarray}
    \Wfin &=& \As\sigma = 4\pi\sigma \Rs^2,
\end{eqnarray}
where $\As$ is the droplet surface area.

The third contribution in Eq.~\eqref{eq:WWbWfWc} is the Coulomb energy associated with the net electric charge of the droplet. Modeling the droplet as a uniformly charged sphere\footnote{This assumption can be relaxed by allowing the electric charges to rearrange and screen the droplet charge~\cite{Iida:1997ay,Iida:1998pi,Heiselberg:1993dc}; see~\cite{Guerrini:2025ojt}. In this work we adopt a different scheme, described below, which yields similar results.}, the Coulomb energy reads (see, e.g.,~\cite{Heiselberg:1992dx,DiToro:2006bkw})
\begin{eqnarray}
    \Wcoul &=&  \frac{16}{15}\,\alphaem\,\pi^2\,
\left[n_B^{\Qs}\left(Y_C^{\Qs}-Y_e^{\Qs}\right)\right]^2\Rs^{5},\label{eq:Wcoul}
\end{eqnarray}
where $\alphaem\simeq 1/137$ is the fine-structure constant and $\dnC\equiv n_B^{\Qs}(Y_C^{\Qs}-Y_e^{\Qs})$ is the net electric charge density of the droplet, i.e. the difference between the charge carried by the quarks and that carried by the leptons trapped inside it. This term was absent in~\cite{Guerrini:2025mxx}, where exact local charge neutrality ($\dnC=0$) was enforced.

The three contributions scale with different powers of $R$ and therefore dominate in different regimes. For small droplets the surface term $\Wfin\propto R^2$ controls the growth of $W$, whereas the bulk term $\Wbulk\propto R^3$ takes over at large $R$. While $\Wbulk$ can have either sign depending on whether the hadronic phase is stable or metastable, $\Wfin$ and $\Wcoul$ are strictly positive and represent the energy cost of forming a finite droplet rather than an infinite bulk region. The Coulomb term $\propto R^5$ would formally dominate at the largest radii, which could preclude the existence of a finite critical droplet; in practice, however, the same electrostatics that generates this term also drives a rearrangement of the leptons that screens the droplet charge and strongly suppresses it, as discussed below and in~\cite{Guerrini:2025ojt,Iida:1997ay}.

Since the nucleation rate depends exponentially on $W$ (see Sec.~\ref{sec:conversion_nucleation_rate}), the transition is overwhelmingly dominated by the least costly path connecting the two phases, i.e. by the saddle point of $W$ in the space of droplet configurations~\cite{Langer:1969bc,Langer:1974cpa}. The saddle-point configuration is a maximum with respect to the droplet radius $R$ and a minimum with respect to the remaining variables ($n_B^{\Qs},\{Y_i^{\Qs}\},\TQs$), while the initial hadronic conditions ($\nBH,\{\YiH\},T$) act as fixed inputs. Carrying out this constrained extremization, described in detail in~\cite{Guerrini:2025mxx,Guerrini:2025ojt}, yields
\begin{eqnarray}
\mu_B^{\Qs} &=& \mu_B^{H}\label{eq:muBminim},\\
\mu_C^{\Qs}+\mu_e^{\Qs} &=& \mu_C^{H}+\mu_e^{H},\label{eq:muCmueminim}\\
\mu_S^{\Qs} &=& \mu_S^{H},\label{eq:muSminim}\\
\mu_e^{\Qs} &=& \mu_e^{H} + \frac{8}{5}\,\pi\,\alphaem\,\Rs^2\,\dnC,\label{eq:muemueminim}\\
\mu_{\nu_e}^{\Qs} &=& \mu_{\nu_e}^{H},\label{eq:munuemunueminim}\\
\TQs &=& T, \label{eq:TQT}
\end{eqnarray}
for the unpaired quark matter case. Equation~\eqref{eq:TQT} expresses thermal equilibrium between the droplet and the surrounding medium, while Eqs.~(\ref{eq:muBminim},\ref{eq:muCmueminim},\ref{eq:muSminim})  enforce chemical equilibrium with respect to the charges conserved by strong interactions (i.e. baryon number $B$, electric charge $C$, and strangeness $S$), see~\cite{Guerrini:2025ojt,Hempel:2009vp}. In the CFL case the composition is locked to $Y_{C}^{Q_{\rm CFL*}}=0$, $Y_{S}^{Q_{\rm CFL*}}=1$, and  $\mu_B^{\Qs}+\mu_S^{\Qs}=\mu_u^{\Qs}+\mu_d^{\Qs}+\mu_s^{\Qs}=\mu_B^{H}+\mu_S^{H}$ replace Eqs.~(\ref{eq:muBminim},\ref{eq:muCmueminim},\ref{eq:muSminim}).

Physically, Eqs.~(\ref{eq:muBminim}--\ref{eq:munuemunueminim}) allow the flavor composition of the droplet to fluctuate locally to the configuration most favorable for nucleation, so that the droplet can acquire a composition different from that of its surroundings while global flavor conservation is preserved. Such an optimized composition may arise from local fluctuations of the initial hadronic composition, from the exchange of strong charges across the phase boundary during droplet formation (including $q\bar q$ pair creation with subsequent charge separation), or from a combination of the two; a detailed discussion is given in~\cite{Guerrini:2025mxx,Guerrini:2025ojt}.

Equation~\eqref{eq:muemueminim} shows that the electron chemical potential inside the droplet is shifted, relative to the value it would have across a sharp interface, by an amount proportional to the net charge density $\dnC$ and to $\Rs^2$. This shift is a direct consequence of minimizing $\Wcoul$: the system pays a price in the bulk free energy in order to reduce the electrostatic energy, which is precisely the effect that a self-consistent charge-screening treatment would produce. Alternative prescriptions for the Coulomb energy of the droplet are compared in~\cite{Guerrini:2025ojt}; see also Appendix~\ref{app:screening}.

Substituting Eqs.~(\ref{eq:muBminim}--\ref{eq:TQT}), or their CFL counterparts, into Eq.~\eqref{eq:WWbWfWc}, the work required to form a droplet of arbitrary radius $R$ reduces to
\begin{eqnarray}
    W(\nBH,\{\YiH\},T,R) &=& -\frac{4}{3}\pi R^3\left(P_{\Qs}-\PH\right)\label{eq:Wfin}\\&&+4\pi \sigma R^2\notag\\&& -\frac{16}{15}\alphaem\pi^2 \dnC^2 R^5,\notag
\end{eqnarray}
with $P_{\Qs}\equiv P_Q(n_B^{\Qs},\{Y_i^{\Qs}\},\TQs)$ evaluated at the composition fixed by Eqs.~(\ref{eq:muBminim}--\ref{eq:TQT}) or their CFL counterparts. It is worth stressing that the last term in Eq.~\eqref{eq:Wfin} is the sum of the (positive) Coulomb energy of Eq.~\eqref{eq:Wcoul} and a bulk contribution generated by inserting the shifted electron chemical potential of Eq.~\eqref{eq:muemueminim} into the bulk term [Eq.~\eqref{eq:Wbulk}], the two combining into a single effective $R^5$ contribution (see~\cite{Guerrini:2025ojt} for the explicit derivation). 

The critical radius $\Rs$ is the value of $R$ that maximizes $W(R)$, with $W(\Rs)\equiv \Ws$. Extremizing Eq.~\eqref{eq:WWbWfWc} at fixed composition and then replacing Eqs.~(\ref{eq:muBminim}--\ref{eq:TQT}) (or their CFL counterparts) gives the condition
\begin{equation}
   P_{\Qs}
   + \frac{4}{15}\,\pi\,\alphaem\,\dnC^2\, \Rs^2
   - \frac{2\sigma}{\Rs}
   \;=\;
   \PH \label{eq:PQPH}
\end{equation}
which expresses the (unstable) mechanical equilibrium of the droplet. Note that, as discussed above for the $\sim R^5$ term of Eq.~\eqref{eq:Wfin}, the $\Rs^2$ term of Eq.~\eqref{eq:PQPH} is not purely electrostatic: it is the sum of the Coulomb contribution and the part of the bulk term that acquires a $\sim \Rs^2$ dependence after replacing Eq.~\eqref{eq:muemueminim}, acting as an effective Coulomb term.

A final clarification concerns the shape of $W(R)$. In the absence of the Coulomb term, $W(R)$ grows monotonically for $R>0$ when the hadronic phase is stable ($\Wbulk>0$), whereas in the metastable case ($\Wbulk<0$) it rises to a maximum at $R=\Rs$ and then decreases indefinitely, defining the usual nucleation barrier. Including the Coulomb term, a metastable hadronic phase may in principle develop no maximum at all (if the Coulomb energy is very large) or a second minimum at large $R$, in which case droplet growth would halt at a configuration less favorable than the initial state and no critical radius would exist in the usual sense. The electron rearrangement encoded in Eq.~\eqref{eq:muemueminim}, however, acts as an effective charge screening and strongly reduces the Coulomb term, so that in all cases considered here $W(R)$ retains the standard barrier shape, with a maximum at $\Rs$ followed by $W(R)<0$ for sufficiently large droplets. Any additional extremum is pushed to $R\gg \Rs$ and does not affect nucleation. In practice, the values of $\Rs$ and $\Ws$ obtained with the present scheme, and with the charge-screening scheme of~\cite{Guerrini:2025ojt}, lie between the two limiting cases of global charge neutrality without Coulomb corrections and strict local charge neutrality. Further details are given in Appendix~\ref{app:screening}.

\subsubsection{Role of color superconductivity}
\label{sec:conversion_nucleation_unpcfl}

In this work we assume that the absolutely stable SQM phase inside QSs is color superconducting. A key issue concerns the behavior of color superconductivity in finite systems~\cite{Amore:2001uf}, such as the first nucleating SQM droplet. Since the diquark coherence length is $\sim 1/\Delta$~\cite{Alford:2007xm}, with $\Delta$ the pairing gap, diquark pairs can only form in droplets larger than this characteristic size. We therefore impose a minimal radius $\Rgap=1/\Delta$, above which the droplet can become superconducting. This scheme was introduced in~\cite{Guerrini:2025mxx,Guerrini:2025ojt}.

For simplicity, we restrict ourselves to the CFL phase. A comprehensive treatment would include other pairing patterns, such as the 2SC phase~\cite{Alford:2007xm}, but our aim here is not to analyze the competition among color-superconducting phases. Rather, we focus on how nucleation is modified when the energetically favored superconducting phase can develop only once $R\geq \Rgap$, while smaller droplets ($R<\Rgap$) remain in the unpaired phase. The same framework can, in principle, be applied to other color-superconducting phases.

Within this approach, the pressure $P_{\Qs}$ entering $W(R)$ in Eq.~\eqref{eq:Wfin} becomes explicitly radius dependent,
\begin{eqnarray}
 P_Q&&(n_B^{\Qs},\{Y_i^{\Qs}\},T,R) =  \label{eq:framPQ}\\
&& \begin{cases}
     P_{\Qunp}(n_B^{\Qs},\{Y_i^{\Qs}\},T), & R \leq \Rgap(T), \\[4pt]
        P_{\Qcfl}(n_B^{\Qs},\{Y_i^{\Qs}\},T), & R > \Rgap(T),
 \end{cases}\notag
\end{eqnarray}
where $\Rgap(T)=1/\Delta(T)$ is set by the temperature-dependent gap of Eq.~\eqref{eq:gap}, and the unpaired ($\Qunp$) and CFL ($\Qcfl$) EOSs are those detailed in Sec.~\ref{sec:eos_quark}. Below $\Rgap$ the droplet is smaller than the diquark coherence length, pairing cannot develop, and only the unpaired phase is available. Above $\Rgap$ pairing is geometrically allowed and the droplet becomes color superconducting.\footnote{Throughout this section the label CFL, when applied to the $R>\Rgap$ branch, denotes the energetically favored phase above the crossover radius $\Rgap$. Because the CFL phase locks the composition to $n_u=n_d=n_s$, for a sufficiently small gap (e.g. when $T \rightarrow \TcCFL$) the pairing gain may not compensate the free-energy cost of this flavor locking, and the unpaired phase remains favored even for $R>\Rgap$; in that case, $\Qcfl$ is to be read as $\Qunp$.} The same switching applies to all other thermodynamic quantities. As $T\to (\TcCFL)^-$ the unpaired region $R\leq \Rgap$ progressively widens, until at $\Delta(T)=0$ (i.e. $T\geq \TcCFL$) one has $\Rgap\to\infty$: the second branch is never activated and the unpaired EOS is recovered at all radii.

Because $P_Q$ switches at $\Rgap$, the work of formation $W(\nBH,\{\YiH\},T,R)$ built from Eq.~\eqref{eq:Wfin} is a piecewise function of $R$: below $\Rgap$ it coincides with the unpaired barrier $\Wunp(R)$, while above $\Rgap$ it follows the CFL barrier $\Wcfl(R)$,
\begin{equation}
    W(R)=
    \begin{cases}
        \Wunp(R), & R \leq \Rgap(T),\\[4pt]
        \Wcfl(R), & R > \Rgap(T),
    \end{cases}
    \label{eq:Wunpcfl}
\end{equation}
where $\Wunp$ ($\Wcfl$) is the barrier obtained using $P_{Q}\to P_{\Qunp}$ ($P_{Q}\to P_{\Qcfl}$) in Eq.~\eqref{eq:Wfin} together with the corresponding composition conditions of Sec.~\ref{sec:conversion_nucleation_energybarrier}.

The critical configuration is the global maximum of $W(R)$,
\begin{equation}
    \Ws=\max_R W(R) \equiv W(\Rs),
    \label{eq:Wstarunpcfl}
\end{equation}
which must be selected by comparing the heights of the candidate maxima. Denoting by $\Rsunp$ and $\Rscfl$ the solutions of Eq.~\eqref{eq:PQPH} for the unpaired and CFL EOSs, respectively, and by $\Wsunp\equiv \Wunp(\Rsunp)$ and $\Wscfl\equiv \Wcfl(\Rscfl)$ the corresponding barrier heights, three scenarios emerge:
\begin{itemize}
    \item if $\Rsunp\leq \Rgap$, the barrier is crossed while the droplet is still unpaired, thus $\Rs=\Rsunp$, $\Ws=\Wsunp$;
    \item if $\Rsunp> \Rgap$ but $\Rscfl\leq \Rgap$, the unpaired branch is still rising at $\Rgap$ while the CFL branch is already decreasing, so the maximum sits at $\Rs=\Rgap$ and $\Ws=\Wunp(\Rgap)$;
    \item if both $\Rsunp> \Rgap$ and $\Rscfl> \Rgap$, the global barrier is $\Ws=\max\!\left[\Wunp(\Rgap),\,\Wscfl\right]$, with $\Rs$ the corresponding radius. The maximum accounts for the possibility that the CFL peak is below the value of $\Wunp$ at the crossover radius $\Rgap$.
\end{itemize}
In the following, this construction is referred to as unpCFL. Unless explicitly stated otherwise, unpCFL is the standard scheme adopted in our results and discussions.

\subsubsection{Nucleation rate and time}
\label{sec:conversion_nucleation_rate}

The rate of a spontaneous thermal fluctuation generating a critical droplet, as a function of the initial hadronic conditions $(\nBH,\{\YiH\},T)$, is~\cite{Landau:1980mil,Langer:1969bc,Langer:1974cpa}
\begin{equation}
    \Gamma=\Gzero \exp\left[-\frac{\Ws}{T}\right],\label{eq:nucreate}
\end{equation}
where $\Ws$ is the energy barrier\footnote{Although we present nucleation here in a thermodynamic formalism, the same expression for the nucleation rate can be derived within a field-theoretic framework; see, e.g.,~\cite{Bessa:2008nw,Mintz:2009ay}.}. While Eq.~\eqref{eq:nucreate} is general, the prefactor $\Gzero$ depends on the specific properties of the system. In the thermal nucleation theory of~\cite{Langer:1969bc,Langer:1974cpa}, $\Gzero$ is expressed as the product of a statistical prefactor $\Omegazero$, measuring the phase-space volume of the saddle-point region, and a dynamical prefactor $\kappa$, the exponential growth rate of the droplet at the saddle point,
\begin{equation}
    \Gzero = \frac{\kappa}{2\pi}\,\Omegazero.
\end{equation}
The quantities $\kappa$ and $\Omegazero$ were computed explicitly in~\cite{Kapusta:2007xjq,Venugopalan:1993vk,Csernai:1992tj}, where nucleation theory was extended to relativistic systems. The statistical prefactor is
\begin{equation}
    \Omegazero=\frac{2}{3\sqrt{3}}\left(\frac{\sigma}{T}\right)^{3/2}\left(\frac{\Rs}{\xi_q}\right)^4,
\end{equation}
with $\xi_q$ the quark correlation length. For the dynamical prefactor we adopt the expression of~\cite{Venugopalan:1993vk},
\begin{equation}
    \kappa = \frac{2\sigma}{\Rs^3\left(\Delta w_*\right)^2}\left[\lambda T+2\left(\frac{4}{3}\eta + \zeta\right)\right],
\end{equation}
where $\Delta w_*=P_{\Qs}+\varepsilon_{\Qs}-P_{H}-\varepsilon_{H}$ is the enthalpy-density difference between the two phases at the saddle point, $\lambda$ is the thermal conductivity, and $\eta$ and $\zeta$ are the shear and bulk viscosities of hadronic matter, respectively. These transport coefficients were estimated in~\cite{Danielewicz:1984kt}, and we adopt the same values as in~\cite{Bombaci:2009jt,Bombaci:2016xuj}. For a comprehensive discussion of the prefactor see~\cite{Kapusta:2007xjq}. We stress, however, that the dominant physics is contained in the exponential, while the prefactor plays only a subleading role; in numerous works the state-of-the-art prefactor is replaced for simplicity by a dimensionally motivated $T^4$ or $\mu^4$, without qualitative impact on the results (see e.g.~\cite{Olesen:1993ek,Mintz:2009ay,Carmo:2013fr} and Sec.~\ref{sec:results_robustness}). 

The nucleation time, i.e. the typical time after which a critical fluctuation is expected to occur in the system, is
\begin{equation}
    \tau=\frac{1}{V\,\Gamma}, \label{eq:taunuc}
\end{equation}
where $V$ is the volume considered. We emphasize that this is neither the lifetime of a single droplet nor the time needed for a single droplet to reach the critical size, but rather the expected time for at least one critical droplet to appear statistically within the system. In astrophysical applications, the system has a space- and time-dependent configuration $\nBH(\mathbf{r},t),\{\YiH(\mathbf{r},t)\},T(\mathbf{r},t)$, and the expected number of nucleation events is
\begin{equation}
    \int_{t_{\rm TOT}} dt \int_{V_{\rm TOT}} d^3\mathbf{r}\;\Gamma\!\left[\nBH(\mathbf{r},t),\{\YiH(\mathbf{r},t)\},T(\mathbf{r},t)\right],
\end{equation}
where $t_{\rm TOT}$ and $V_{\rm TOT}$ denote the total time and volume considered. In numerical simulations of BCSMs, CCSNe, or PNSs, the system is discretized into cells of volume $V$ (typically $\sim 100$~m of linear size), within which $(\nBH,\{\YiH\},T)$ can be regarded as approximately constant over both space and time during a time step $\Delta t$. The local nucleation rate can then be evaluated cell by cell using the simulation output as input for $\Gamma$, and the phase transition is statistically triggered in cell $r$ at time $t$ if
\begin{equation}
    \Delta t\, V_{t,r}\, \Gamma\!\left[\nBH,\{\YiH\},T\right]_{t,r} \gtrsim  1.
\end{equation}

\subsubsection{Nucleation in a proto-neutron star}
\label{sec:conversion_nucleation_pns}

In this work we do not have access to an astrophysical simulation providing the input $(\nBH,\{\YiH\},T)$. Since the evolution of a hadronic PNS is slower than the weak-equilibration timescale, we can assume that the hadronic PNS is in chemical equilibrium, so that $(\nBH,\{\YiH\},T)\to(\nBH,\YLH,T)$. Our strategy is to compute the thermodynamic conditions $(\nBH,\YLH,T)$ under which a nucleation event is statistically expected to occur in a hadronic PNS before these conditions change, i.e. when the nucleation time $\tau$ of Eq.~\eqref{eq:taunuc} is shorter than the dynamical timescale of the system, $\taudyn\simeq 10^{-3}$~s.

In principle, one should compute the local nucleation rate throughout the entire PNS and then integrate over the volume. However, since the nucleation rate depends exponentially on $\Ws/T$, the integral is dominated by the region where $\Ws/T$ lies close to its minimum, i.e.\ the stellar center. For this reason, we evaluate $\Gamma$ at the central conditions and adopt as reference volume a sphere of radius $100$~m, i.e. $V^{\rm PNS}_{\rm center}=\tfrac{4}{3}\pi(100~{\rm m})^3\simeq 4\times10^{51}~{\rm fm^3}$, as in~\cite{Bombaci:2016xuj}, and comparable to the cell size of astrophysical simulations (see Sec.~\ref{sec:conversion_nucleation_rate}). This volume is a convention rather than the size of the nucleating region (see Sec.~\ref{sec:results_robustness}). Similarly, during the PNS evolution the central conditions $(\nBH,\YLH,T)$ can remain approximately constant on timescales longer than $\taudyn$. Both simplifications can only underestimate the rate, so our approach may slightly overestimate the values of $\nBH$ and $T$ at which nucleation occurs. Nevertheless, as shown in Sec.~\ref{sec:results} (see also~\cite{Bombaci:2016xuj,Guerrini:2024gzu}), $\tau$ varies extremely rapidly with the thermodynamic conditions, so adopting a reference timescale different from $\taudyn$, or a different reference volume, does not significantly affect the results (see Sec.~\ref{sec:results_robustness}). We therefore define the \emph{nucleation conditions} as the set of parameters $(\nBH,\YLH,T)$ such that
\begin{equation}
    \tau \equiv \left[V^{\rm PNS}_{\rm center}\,\Gamma(\nBH,\YLH,T)\right]^{-1} \leq \taudyn.
\end{equation}

Since we cannot use dedicated PNS simulations for our EOS model, we discuss the time evolution of the PNS central conditions by considering two representative snapshots, with typical thermodynamic values extrapolated from existing simulations~\cite{Pons:1998mm,Pons:2001ar,Camelio:2017nka,Camelio:2017hrs,Roberts:2012um}:
\begin{itemize}
\item ($\tzero$) $\SH/N_B=1.5$ and $\YLH=0.35$ at the time of the CCSN explosion;
\item ($\tTmax$) $\SH/N_B=2$ and $\YLH=0.25$ after $\sim 10$~s, when the central temperature reaches its maximum,
\end{itemize}
where $\SH/N_B=s_H/\nBH$ denotes the entropy per baryon of the hadronic phase. In computing the PNS masses and radii we therefore assume that the star is characterized by fixed values of $\SH/N_B$ and $\YLH$, and the corresponding central baryon density is fixed at given $\SH/N_B,\YLH$ (and thus for $\tzero$ and $\tTmax$) as a function of the PNS mass.

\subsection{Remnant and energy released}
\label{sec:conversion_nucleation_remnant}
Once the nucleation conditions are reached in a hadronic PNS, the seed grows and a conversion front propagates outward, transforming hadronic matter into SQM. The remnant is a hot QS that subsequently cools and deleptonizes via intense neutrino emission until it settles into a hydrostatic configuration on the QS branch~\cite{Pagliara:2013tza,Drago:2015fpa,Drago:2015dea}.

During the conversion, both strong and weak processes in the front conserve the total baryon number, while neutrinos and gravitational waves carry away energy and lepton number but not baryons. Significant deviations would require substantial mass ejection across the surface, which is unlikely because a PNS lacks an extended envelope to ablate. Thus, we assume that the total baryon number $N_B^{\rm PNS}$ is conserved during the conversion, so that the final QS has the same baryonic mass ($\MB^{\rm PNS} \equiv m_b\,N_B^{\rm PNS}$)\footnote{We fix the baryon mass to the atomic mass unit $m_b=931$ MeV. Other works sometimes adopt $m_b=939$ MeV. This choice is arbitrary as long as it is applied consistently to all objects.} as the initial PNS, i.e. $\MB^{\rm QS}=\MB^{\rm PNS}$~\cite{Bombaci:2000cv,Bombaci:2016xuj}.

By contrast, the gravitational mass of the star changes during the conversion. The total released energy is given by the difference in binding energies between the final QS and the initial PNS:
\begin{equation}
    \Econv = {\rm B.E.}_{\rm QS} - {\rm B.E.}_{\rm PNS} = \MPNS-\MQS.
\end{equation}
This energy can be released as a powerful neutrino burst or intense gravitational waves, and could possibly even power gamma-ray bursts (GRBs)~\cite{Berezhiani:2002ks,Berezhiani:2002ea,Drago:2015qwa,Drago:2015dea}. 
Depending on the stage of the PNS evolution at which the conversion occurs, and on the timescale required to complete it, such a GRB could be delayed with respect to the CCSN explosion. 
In addition, a delayed conversion could impart a ``second kick'' to the QS, providing a natural explanation for the bimodal distribution of pulsar velocities~\cite{Bombaci:2004nu}. 
Finally, if the converted QS resides in a binary system, the nearly instantaneous\footnote{The conversion timescale is much shorter than a typical orbital period.} gravitational mass defect can increase the orbital eccentricity, even in systems previously circularized by tides (see~\cite{Chanlaridis:2024rov} for a recent study in the twin-star scenario). 

On the other hand, if the nucleation conditions are never reached during the evolution of the PNS, the remnant will instead cool and settle into a cold NS with the same baryon mass, $\MB^{\rm NS}=\MB^{\rm PNS}$.

\section{Results}
\label{sec:results}
We now apply the nucleation formalism developed in Sec.~\ref{sec:conversion} to study the process of deconfinement of hadronic PNSs, discussing its consequences for the coexistence of NSs and QSs within the two-families scenario. First, we present the macroscopic and thermodynamic properties of the adopted EOSs and their corresponding stellar sequences. Next, we investigate the role of color superconductivity in the nucleation of a critical droplet. Finally, we explore the implications for the PNS evolution, focusing on the parameter space that permits the coexistence of NSs and QSs.

\subsection{Compact star characteristics}
\label{sec:results_compactstars}

\begin{figure*}
    \centering
    \includegraphics[width=0.9\textwidth]{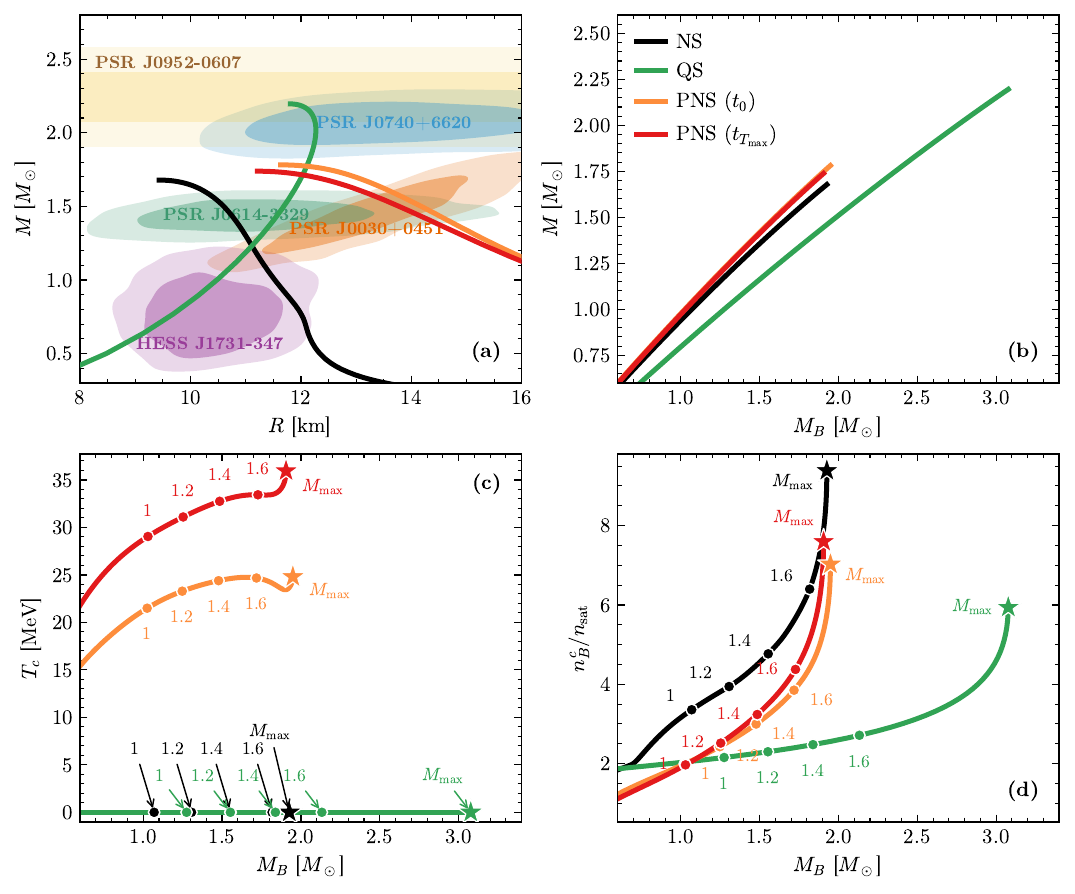}
    \caption{Macroscopic and thermodynamic characteristics of the stellar configurations. (a) Mass--radius relations for cold NSs (black), QSs (green), and hot PNSs at the two snapshots $\tzero$ (orange) and $\tTmax$ (red), compared with some recent observational constraints~\cite{Romani:2022jhd,Miller:2021qha,Miller:2019cac,Mauviard:2025dmd,Doroshenko:2022nwp}. (b) Gravitational mass $M$ versus baryonic mass $\MB$ for the same stellar sequences. (c) Central temperature $\Tcen$ and (d) central baryon density $\nBc$ as functions of $\MB$, with dots indicating $\MPNS=(1.0,1.2,1.4,1.6)\,\Msol$ and stars denoting $\Mmax$. Since PNS evolution can be considered at constant $\MB$, it can be tracked along vertical lines in panels~(b), (c), and~(d). The SQM parameters are $\Bq=145\MeV$, $\Dzero=80\MeV$, $\alphas=\tfrac{\pi}{2}\times0.1$, and $\ms=100\MeV$.}
    \label{fig:seq}
\end{figure*}

Figure~\ref{fig:seq} summarizes the macroscopic and thermodynamic properties of the stellar configurations. The black and green curves represent, respectively, the cold, neutrinoless, weak-equilibrium hadronic NS branch, computed with the SFHo-H$\Delta$ EOS, and QS branch, obtained with a color-superconducting quark-matter EOS (see Sec.~\ref{sec:eos}). These two branches correspond to the two possible final outcomes of PNS evolution, i.e. the metastable NS and the stable QS configurations that define the two-families scenario. For this figure we adopt the SQM parametrization $\Bq=145\MeV$, $\Dzero=80\MeV$, $\alphas=\tfrac{\pi}{2}\times0.1$, and $\ms=100\MeV$ (``Set~A'' in what follows); we use it for most figures in this section, unless stated otherwise, and present a comprehensive analysis of the broader SQM parameter space in Sec.~\ref{sec:results_paramspace}.

Panel~(a) shows the mass--radius relations. For the adopted parametrization, the two families reach maximum gravitational masses $\MmaxNS\simeq1.68\,\Msol$ (at central baryon density $\nBc\simeq9.4\,\nsat$ and radius $9.4$~km) and $\MmaxQS\simeq2.20\,\Msol$ (at $\nBc\simeq5.9\,\nsat$ and $11.8$~km), while a canonical $1.4\,\Msol$ star has a radius $10.8$~km on the NS branch. We also display the two representative hot PNS sequences used throughout this work, as discussed in Sec.~\ref{sec:conversion_nucleation_pns}: the configuration at the time of the CCSN explosion ($\tzero$: $\SH/N_B=1.5$, $\YLH=0.35$, orange curve) and at the epoch of maximal central temperature ($\tTmax$: $\SH/N_B=2$, $\YLH=0.25$, red curve). We stress that, within the two-families scenario, a hadronic EOS unable to support $\sim2\Msol$ is not in conflict with observations of massive pulsars, since the most massive objects are identified with QSs~\cite{Passarella:2026mxy}.

Panel~(b) displays the gravitational mass $M$ as a function of the baryonic mass $\MB$ for all four sequences. Once the PNS baryonic mass is fixed at explosion, the star evolves at nearly constant $\MB$, so that its secular evolution proceeds vertically in this diagram (and in panels~c and~d) from the hot snapshots down to the cold NS or QS branch, depending on whether deconfinement occurs (see Sec.~\ref{sec:conversion_nucleation_remnant}). For the hadronic EOS adopted here, $\MBmax^{\rm PNS}(\tTmax)\simeq1.91\Msol<\MBmax^{\rm NS}\simeq1.93\Msol<\MBmax^{\rm PNS}(\tzero)\simeq1.95\Msol$, while the QS branch extends to significantly larger baryonic masses ($\MBmax^{\rm QS}\simeq3.1\,\Msol$), reflecting the larger binding of SQM. Since $\MBmax^{\rm PNS}$ decreases between $\tzero$ and $\tTmax$, a PNS born at $\tzero$ with $\MB^{\rm PNS}>\MBmax^{\rm PNS}(\tTmax)$ will exceed the maximum supportable baryonic mass as it heats and deleptonizes, and therefore collapses into a BH before reaching $\tTmax$, unless nucleation is triggered first. Conversely, any PNS that does reach $\tTmax$ has a stable, cold NS counterpart at the same $\MB$ and cools into it if deconfinement does not occur. This ordering is in general model dependent: if $\MBmax^{\rm PNS}(\tTmax)$ exceeded $\MBmax^{\rm NS}$, a PNS with $\MB^{\rm PNS}$ in between would survive $\tTmax$ but collapse into a BH during cooling, unless it converted into a QS first.

The gravitational-baryonic mass relation also constrains where PNS conversion populates the QS branch. If nucleation is attained in the maximum-mass PNS configuration at $\tTmax$ ($\MmaxPNS(\tTmax)=1.74\,\Msol$), the resulting QS has mass $\simeq1.45\Msol$. Hence the conversion of hadronic PNSs does not populate the high-mass end of the QS branch; the most massive QSs must instead arise from other evolutionary channels, such as BCSMs, fallback accretion between core collapse and explosion of very massive progenitors, or later mass accretion from a binary companion.

Panels~(c) and~(d) display the central temperature $\Tcen$ and the central baryon density $\nBc$ along the PNS sequences as functions of $\MB$, with markers at $\MPNS=(1.0,1.2,1.4,1.6)\,\Msol$ and at $\Mmax$. The central temperature increases with mass and is markedly higher at $\tTmax$ (reaching $\Tcen\simeq36\MeV$ near $\Mmax$) than at $\tzero$ ($\Tcen\simeq25\MeV$); by definition, the cold NS and QS sequences lie at $\Tcen=0$. The central density likewise increases steeply with mass and, at fixed $\MB$, is largest on the cold NS branch. These trends foreshadow the nucleation analysis: heavier PNSs reach both the higher densities and the higher temperatures that strongly favor deconfinement.

\subsection{Impact of finite-size pairing on the critical droplet}
\label{sec:results_colorsup}

\begin{figure*}
    \centering
    \includegraphics[width=0.9\textwidth]{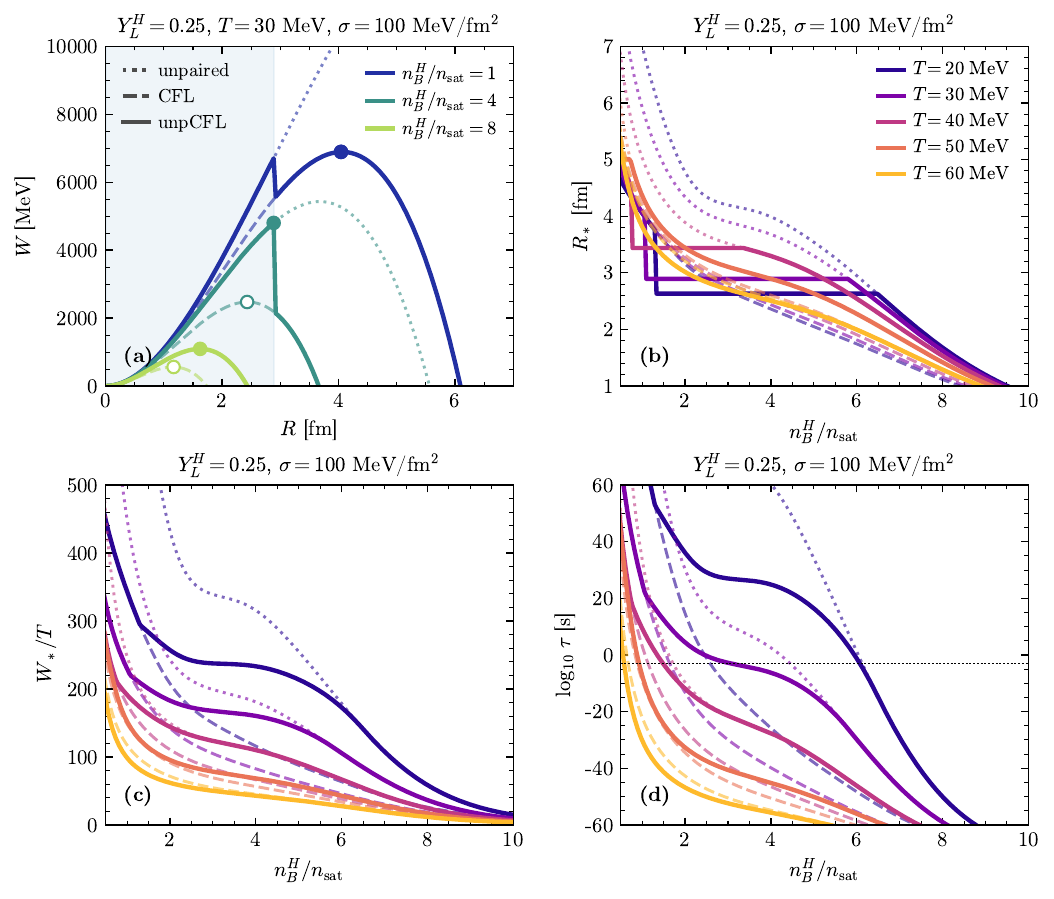}
    \caption{Nucleation of a critical droplet under three prescriptions for the droplet EOS: unpaired (dotted), CFL (dashed), and the unpCFL scheme (solid).
    (a) Work of formation $W(R)$ at $T=30\MeV$ for $\nBH/\nsat=1,4,8$; filled/open circles mark the unpCFL/CFL critical points, and the shaded region indicates $R<\Rgap$, where pairing is geometrically forbidden.
    (b) Critical radius $\Rs$,
    (c) reduced barrier $\Ws/T$, and
    (d) $\log_{10}\tau$ [s] (for $V=V^{\rm PNS}_{\rm center}$) versus $\nBH/\nsat$ for $T=20$--$60\MeV$;
    the dotted horizontal line in~(d) marks $\taudyn\simeq10^{-3}$~s.
    Here $\YLH=0.25$ and $\sigma=100\MeVfm$. }
    \label{fig:barrier}
\end{figure*}

Before turning to the astrophysical application, we focus now on the impact of the finite-size treatment of color superconductivity introduced in Sec.~\ref{sec:conversion_nucleation_unpcfl}. Figure~\ref{fig:barrier} shows, for a hadronic phase at fixed lepton fraction $\YLH=0.25$ and surface tension $\sigma=100\MeVfm$, the nucleation barrier and the derived critical quantities under three prescriptions for the droplet EOS: the purely \emph{unpaired} phase (dotted), the purely \emph{CFL} phase (dashed), and the physically motivated \emph{unpCFL} scheme (solid), in which the droplet is in the unpaired phase below the radius $\Rgap=1/\Delta$ and in the CFL phase above it.

Panel~(a) shows the work of formation $W(R)$ of a droplet at $T=30\MeV$ for three representative hadronic densities. In the unpaired case the barrier is high and the critical radius large, because unpaired quark matter is comparatively less bound; in the pure-CFL case pairing lowers the bulk free energy, reducing both the barrier and the critical radius. The unpCFL work of formation follows the unpaired branch for $R<\Rgap$ and the CFL branch for $R\ge \Rgap$, with a downward jump at $R=\Rgap$ where the (lower) CFL free energy switches on.

The critical configuration of the unpCFL scheme is the global maximum of this piecewise barrier, and its location relative to $\Rgap$ defines three regimes. Writing $\Rsunp$ ($\Rscfl$) and $\Wsunp\equiv \Wunp(\Rsunp)$ ($\Wscfl\equiv \Wcfl(\Rscfl)$) for the critical radius and barrier of the purely unpaired (CFL) branch, the outcome is fixed by two comparisons: whether $\Rsunp$ lies below or above $\Rgap$, and, in the latter case, whether the cusp value $\Wunp(\Rgap)$ exceeds the CFL peak $\Wscfl$:

\begin{itemize}
  \item[(i)] \emph{high density}: $\Rsunp\le \Rgap$: the unpaired barrier peaks before pairing can set in, so the critical droplet is unpaired, $\Rs=\Rsunp$ and $\Ws=\Wsunp$, and the CFL condensate is irrelevant to nucleation;
  \item[(ii)] \emph{intermediate density}: $\Rsunp>\Rgap$ and $\Wunp(\Rgap)>\Wscfl$: the highest point along the barrier path is the cusp at $\Rgap$, so the critical seed is gap-locked, $\Rs=\Rgap$ and $\Ws=\Wunp(\Rgap)$;
  \item[(iii)] \emph{low density}: $\Rsunp>\Rgap$ and $\Wunp(\Rgap)<\Wscfl$: the CFL peak dominates, so $\Rs=\Rscfl$ and $\Ws=\Wscfl$.
\end{itemize}
Regime~(ii) does not require $\Rscfl<\Rgap$: the same gap-locked critical point is obtained whenever the cusp $\Wunp(\Rgap)$ is the global maximum, whether or not the CFL branch would peak beyond $\Rgap$. In Fig.~\ref{fig:barrier}(a), the case $\nBH/\nsat=1$ realizes regime~(iii) (and sits near the (ii)--(iii) boundary), $\nBH/\nsat=4$ regime~(ii), and $\nBH/\nsat=8$ regime~(i).

Panels~(b)--(d) show, as functions of the hadronic baryon density $\nBH/\nsat$ and for several temperatures, the critical radius $\Rs$, the reduced barrier $\Ws/T$ (the argument of the exponential in the thermal nucleation rate), and the nucleation time $\tau$. 
The three regimes are directly visible: at high (low) density the unpCFL critical quantities coincide with the purely unpaired (CFL) ones, while over an extended intermediate interval the critical radius is pinned at $\Rs=\Rgap$ (the flat plateaus in panel~(b), e.g.\ $\Rs\simeq2.9$~fm at $T=30\MeV$). In this gap-locked regime (ii) the seed size is set by the temperature-dependent gap rather than by the bulk thermodynamics, and the unpCFL reduced barrier and nucleation time lie systematically between the unpaired and CFL curves.

Temperature affects the critical quantities in two opposing ways. The higher the temperature, the larger the bulk-energy gain of unpaired SQM relative to hadronic matter, so $\Rsunp$ decreases; at the same time the gap $\Delta(T)$ shrinks, so $\Rgap(T)=1/\Delta(T)$ grows. As a result, at higher $T$, regime~(ii) enters at lower $\nBH$ and is characterized by a larger gap-locked radius $\Rs=\Rgap$, giving $\Rs$ a non-monotonic dependence on $T$. For $T>\TcCFL$ (e.g.\ the $T=60\MeV$ curve, since $\TcCFL\simeq57\MeV$ for the parametrization used here) the gap vanishes, $\Rgap\to\infty$, and the system is always in regime~(i).

Note that $\Rs$ is discontinuous at the (ii)--(iii) boundary: as $\nBH$ increases, $\Ws$ passes continuously from $\Wscfl$ to $\Wunp(\Rgap)$, whereas $\Rs$ jumps from $\Rscfl$ to $\Rgap$. Since it is $\Ws$ (and not $\Rs$) that varies continuously, this discontinuity appears only in panel~(b) and is absent in panels~(c) and~(d).

These differences are not merely quantitative. The unpCFL nucleation time can exceed the pure-CFL estimate by many orders of magnitude, so neglecting the finite-size suppression of pairing would qualitatively misrepresent the nucleation conditions. 
The horizontal line in panel~(d) marks the dynamical timescale $\taudyn\simeq10^{-3}$~s; the density at which each curve crosses it defines the nucleation threshold at that temperature. The steepness of $\tau(\nBH)$ (tens of orders of magnitude over a factor of two in density, to be compared with the age of the Universe, $\sim10^{17}$~s) is what makes the precise value of the reference timescale and of the reference volume irrelevant for defining the nucleation conditions, as anticipated in Sec.~\ref{sec:conversion_nucleation_pns} and discussed in Sec.~\ref{sec:results_nucleationPNS}.

\subsection{Nucleation conditions in PNSs}
\label{sec:results_nucleationPNS}

\begin{figure*}
    \centering
    \includegraphics[width=0.9\textwidth]{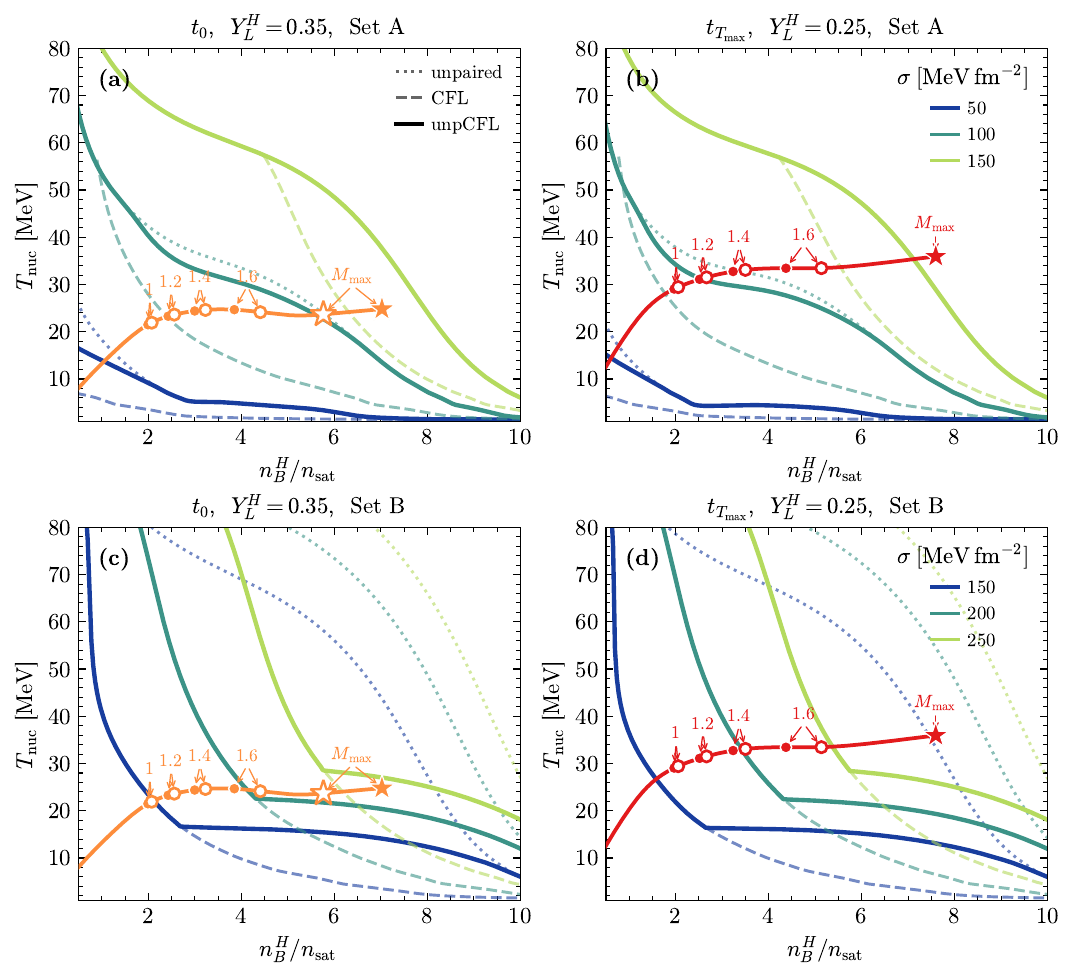}
    \caption{Nucleation temperature $\Tnuc$ versus initial hadronic baryon density $\nBH/\nsat$ for the unpaired (dotted), CFL (dashed), and unpCFL (solid) prescriptions and several surface tensions, for two SQM parametrizations: Set~A ($\Bq=145\MeV$, $\Dzero=80\MeV$, $\alphas=\tfrac{\pi}{2}\times0.1$; top) and Set~B ($\Bq=158\MeV$, $\Dzero=157\MeV$, $\alphas=\tfrac{\pi}{2}\times0.08$; bottom). Orange and red lines show the central-condition sequences of PNSs at $\tzero$ and $\tTmax$, respectively; filled dots label the PNS by its own gravitational mass and open dots by that of the cold NS of equal baryonic mass; the filled star marks the maximum-mass PNS and the open star the PNS with the baryonic mass of the maximum-mass cold NS, which exists only at $\tzero$ since $\MBmax^{\rm PNS}(\tTmax)<\MBmax^{\rm NS}$. A PNS deconfines if its central point lies above the corresponding $\Tnuc$ curve.}
    \label{fig:Tnuc}
\end{figure*}

\begin{figure*}
    \centering
    \includegraphics[width=0.9\textwidth]{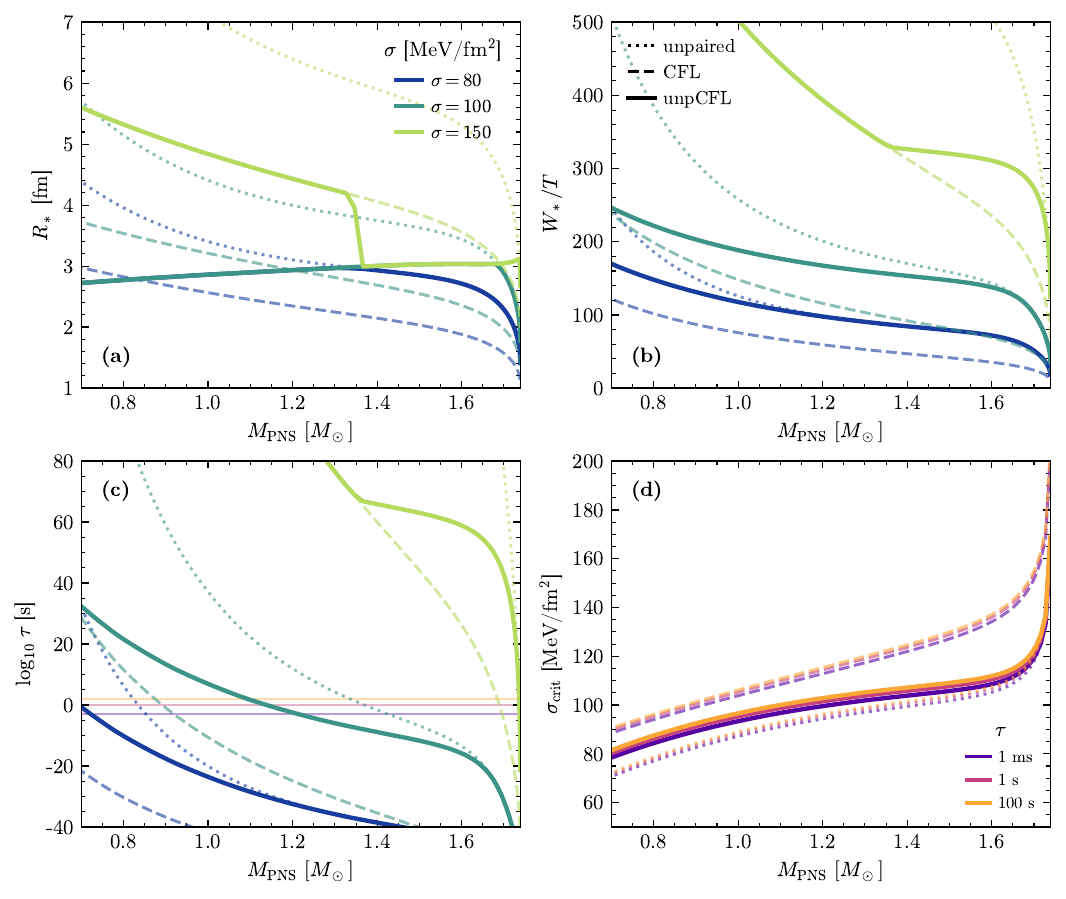}
    \caption{Nucleation at the center of a PNS as a function of gravitational mass $\MPNS$ for snapshot $\tTmax$ and parameter set A.
    (a) Critical radius $\Rs$,
    (b) reduced barrier $\Ws/T$, and
    (c) $\log_{10}\tau$ [s] (for $V=V^{\rm PNS}_{\rm center}$) for $\sigma=(80,100,150)\MeVfm$;
    dotted, dashed, and solid curves denote the unpaired, CFL, and unpCFL prescriptions.
    (d) Critical surface tension $\sigcrit$ versus $\MPNS$ for reference timescales $\tau=(1~{\rm ms},1~{\rm s},100~{\rm s})$; the near-coincidence of the three curves demonstrates the insensitivity of $\sigcrit$ to $\tau$. A PNS deconfines if $\sigma<\sigcrit(\MPNS)$.}
    \label{fig:sigcrit_mass}
\end{figure*}

We now put together the nucleation quantities computed in Sec.~\ref{sec:results_colorsup} with the PNS thermodynamic conditions of Sec.~\ref{sec:results_compactstars}, through Figs.~\ref{fig:Tnuc} and~\ref{fig:sigcrit_mass}, which show the same information from two complementary points of view: the first works in the thermodynamic plane $(\nBH,T)$, the second along the stellar sequence, as a function of $\MPNS$. 

Figure~\ref{fig:Tnuc} shows the \emph{nucleation temperature} $\Tnuc(\nBH,\YLH)$, namely the temperature at which $\tau=\taudyn$ for a given initial hadronic baryon density and lepton fraction, for a range of surface tensions and two SQM parametrizations: a lower-bag/lower-gap set already used in Figs.~\ref{fig:seq} and~\ref{fig:barrier} (``Set~A'': $\Bq=145\MeV$, $\Dzero=80\MeV$, $\alphas=\tfrac{\pi}{2}\times0.1$; panels~a,b) and a higher-bag/higher-gap set (``Set~B'': $\Bq=158\MeV$, $\Dzero=157\MeV$, $\alphas=\tfrac{\pi}{2}\times0.08$; panels~c,d). Superimposed are the sequences of central conditions $(\nBcH,\Tcen)$ of the PNSs at the two snapshots $\tzero$ (orange; panels~a,c) and $\tTmax$ (red; panels~b,d). Along each sequence, filled dots mark the PNS of the labeled gravitational mass, whereas open dots mark the PNS whose baryonic mass equals that of a cold NS of the labeled gravitational mass,  i.e.\ the NS into which the PNS would evolve if nucleation did not occur. A given PNS deconfines when its central point lies above the corresponding $\Tnuc$ curve.\footnote{If $\tau$ were not monotonically decreasing with $\nBH$ and $T$ and crossed $\taudyn$ more than once, we take the crossing at the smallest $T$.} For Set~A and $\sigma=100\MeVfm$, for instance, $\Tnuc$ intersects the $\tTmax$ sequence at the central density of a $\MPNS(\tTmax)\simeq1.21\Msol$ star: lighter PNSs cool and deleptonize into NSs of equal baryonic mass, whereas heavier ones nucleate and convert into QSs.

Each panel displays the three droplet prescriptions of Sec.~\ref{sec:results_colorsup}: the unpaired curve (dotted) lies highest, the CFL curve (dashed) lowest, and the unpCFL curve (solid) interpolates between them, tracing out the regimes~(i)--(iii). Raising $\sigma$ increases the critical work, so that $\tau=\taudyn$ is met only at higher $T$ (at fixed $\nBH$) or higher $\nBH$ (at fixed $T$); the $\Tnuc$ curves therefore shift upward and intersect the PNS sequences at higher density, raising the minimum PNS mass that can nucleate and convert to a QS.

For Set~A (panels~a,b) the small gap keeps the radius $\Rgap=1/\Delta$ large, and this is reinforced at low $\nBH$, where nucleation requires high temperatures that suppress the gap further (indeed $\Delta$ vanishes once $\Tnuc>\TcCFL\simeq57\MeV$). For $\sigma=150\MeVfm$ regime~(i) is therefore realized almost everywhere: it holds at low $\nBH$ because $\Rgap$ is large, and at high $\nBH$ because the barrier, and hence $\Rsunp$, is small enough that the unpaired droplet reaches criticality before pairing can set in at $\Rgap$. A window of regime~(ii) opens only for small enough surface tension ($\sigma\simeq100\MeVfm$) and intermediate density. Regime~(iii) is never realized along the $\Tnuc$ curves of Set~A: where it could occur, at low $\nBH$, nucleation requires temperatures at which the gap is zero or negligible, making unpaired quark matter favored over CFL.

For Set~B (panels~c,d) the large gap pushes $\TcCFL$ above the plotted range and shrinks $\Rgap$, so pairing operates even in small droplets. At low $\nBH$ the CFL peak dominates and nucleation proceeds in regime~(iii); at intermediate $\nBH$ the system crosses into regime~(ii), where the gap-locked cusp at $\Rgap$ exceeds the CFL barrier and therefore retards nucleation relative to the pure-CFL case, and it remains there up to the highest densities shown; regime~(i) could be reached only beyond this range. 

As anticipated in Sec.~\ref{sec:results_compactstars}, the $\tTmax$ snapshot (right panels) is the most favorable to nucleation: at fixed $\sigma$ the PNS sequence meets the $\Tnuc$ curves at lower $\nBH$, lowering the threshold PNS mass for conversion to a QS. Indeed, two independent effects act in the same direction here. The first is the temperature itself: the star is hotter at $\tTmax$, so its central point sits higher in the $(\nBH,T)$ plane. The second is deleptonization: since $\Tnuc$ depends on $\YLH$ as well as on $\nBH$, the whole $\Tnuc$ curve moves down when the lepton fraction drops: at $\sigma=100\MeVfm$ and $\nBH=2$--$5\,\nsat$, going from $\YLH=0.35$ to $0.25$, $\Tnuc$ is reduced by $1$--$6\MeV$.

After $\tTmax$ the star cools and two effects compete. On the one hand, the central density keeps rising toward its cold-branch value [Fig.~\ref{fig:seq}(d)] and the lepton fraction keeps falling; both lower $\Tnuc$ and favor nucleation. On the other hand, the central temperature itself drops. The second effect wins: during cooling $\Tcen$ falls by tens of MeV, whereas the rise of $\nBcH$ and the drop of $\YLH$ lower $\Tnuc$ by much less, so the central point moves away from the nucleation region.

The $\tTmax$ epoch is thus, to a good approximation, the bottleneck for the fate of the star: a PNS that has not nucleated by $\tTmax$ will cool into a NS. For this reason every quantity in the rest of this section, and in particular the critical surface tension introduced below, is evaluated at the $\tTmax$ snapshot.

Figure~\ref{fig:sigcrit_mass} takes the second point of view. Nothing new enters the calculation at this stage: only the independent variable changes, from the thermodynamic state to the mass of the star that realizes it. For each PNS gravitational mass $\MPNS$ the central conditions $(\nBcH,\YLH,\Tcen)$ are fixed by the snapshot (Fig.~\ref{fig:seq}c,d), and the critical droplet properties at given conditions can be evaluated as in Fig.~\ref{fig:barrier}(b)--(d). 
In particular, Fig.~\ref{fig:sigcrit_mass}(a)--(c) shows the critical radius, reduced barrier, and nucleation time at the center of a PNS in the snapshot $\tTmax$, using Set~A, as functions of $\MPNS$, for three values of the surface tension $\sigma=(80,100,150)\MeVfm$ and for the three prescriptions for the droplet EOS (unpaired, CFL, and unpCFL).
Note that the regime structure of Sec.~\ref{sec:results_colorsup} survives the change of variable, and that panels~(a) and~(b) locate the regimes along the stellar sequence: $\Rs$ jumps at the (iii)$\to$(ii) transition, and above it the gap-locked radius $\Rs=\Rgap$ is independent of $\sigma$ and grows with $\MPNS$, because $\Tcen$ rises with $\MPNS$ and the gap $\Delta(T)$ falls accordingly.

The reduced barrier also responds to the surface tension differently in regimes~(i) and~(iii) than in regime~(ii). In the purely unpaired and purely CFL cases, and neglecting the Coulomb term, Eq.~\eqref{eq:PQPH} gives $\Rs=2\sigma/(P_{\Qs}-\PH)$ and hence $\Ws=16\pi\sigma^3/[3(P_{\Qs}-\PH)^2]\propto\sigma^3$, the driving force $P_{\Qs}-\PH$ being independent of $\sigma$. In the gap-locked regime~(ii), by contrast, $\Rs$ is pinned at $\Rgap$, which does not grow with $\sigma$: every term of $W$ is then evaluated at a fixed radius and $\Ws$ becomes \emph{linear} in the surface tension. The Coulomb term modifies both scalings, because through Eq.~\eqref{eq:muemueminim} it makes the droplet composition depend on $\Rs$, and hence on $\sigma$; the effective exponents are therefore somewhat larger than the $\sigma^3$ and $\sigma^1$ quoted above.

The quantity that decides the fate of the star is the nucleation time of panel~(c). As expected, heavier stars have lower barriers and dramatically shorter nucleation times, so that for a given $\sigma$ there is a threshold mass above which nucleation is triggered within $\taudyn$; increasing $\sigma$ raises the barrier uniformly and pushes this threshold to higher mass\footnote{Note that this threshold mass is the one already read off Fig.~\ref{fig:Tnuc}: for $\sigma=100\MeVfm$ the unpCFL curve of Fig.~\ref{fig:sigcrit_mass}(c) crosses $\taudyn$ at $\MPNS\simeq1.21\Msol$, the same mass at which the $\Tnuc$ curve meets the $\tTmax$ sequence for Set~A [Fig.~\ref{fig:Tnuc}(b)].}.

This motivates introducing the notion of \emph{critical surface tension} $\sigcrit(\MPNS)$, defined as the value of $\sigma$ for which the nucleation time at the center of the star, at the $\tTmax$ snapshot, equals a reference timescale $\tau$. A PNS of mass $\MPNS$ nucleates a critical droplet, triggering conversion into a QS, if
\begin{equation}
  \sigma < \sigcrit(\MPNS) ,
  \label{eq:sigcrit_criterion}
\end{equation}
i.e.\ if the physical surface tension is small enough that the barrier can be statistically overcome at the central conditions of the given PNS within a given timescale. The criterion~\eqref{eq:sigcrit_criterion} reduces the question of NS--QS coexistence to the comparison between the (unknown) physical surface tension and a computed, mass-dependent threshold.

Panel~(d) shows $\sigcrit$ as a function of $\MPNS$ for three reference timescales, $\tau=(1~{\rm ms},1~{\rm s},100~{\rm s})$. Two results stand out. First, $\sigcrit$ increases monotonically with mass, from $\simeq93\MeVfm$ at $1\Msol$ to $\simeq119\MeVfm$ at $1.7\Msol$, and then steeply as $\MmaxPNS$ is approached, for the SQM parametrization considered in the figure. Second, $\sigcrit$ is remarkably insensitive to the reference timescale: varying $\tau$ over five orders of magnitude, from $1$~ms (the typical astrophysical dynamical timescale) to $100$~s (a typical upper bound on the duration of the hot PNS phase), shifts $\sigcrit$ at $\MPNS=1.4\Msol$ only from $\simeq104$ to $\simeq 107\MeVfm$, a negligible effect against the other large uncertainties of the problem, such as the EOS itself and the actual behavior of pairing in small droplets. This is the quantitative counterpart of the steep $\tau(\nBH)$ dependence noted above, and it justifies \emph{a posteriori} the use of a single dynamical timescale in defining the \emph{nucleation conditions} as the thermodynamic conditions at which $\tau=\taudyn=1$~ms.

To better understand why one can introduce a representative timescale to fix the nucleation conditions, let us consider again the dependence of $\tau$ on the rate prefactor and on the energy barrier.

Since the nucleation time is controlled by the exponential $\tau^{-1}\propto\exp(-\Ws/T)$, imposing $\tau=\taudyn$ approximately fixes the \emph{reduced barrier} itself. More precisely, Eqs.~\eqref{eq:nucreate} and~\eqref{eq:taunuc} give $\Ws/T=\ln\!\left(V\kappa\Omegazero\tau/2\pi\right)$ exactly, so the required reduced barrier depends on the thermodynamic conditions and on the SQM parameters only logarithmically, through the prefactor. This is what panel~(b) records: the value of $\Ws/T$ reached where the corresponding curve of panel~(c) crosses $\taudyn$ is nearly the same for every surface tension and every prescription. For every combination of droplet prescription and surface tension whose threshold falls inside the mass range of Fig.~\ref{fig:sigcrit_mass}, the nucleation condition is met at $\Ws/T=166\pm2$; its distribution over the whole parameter space, and how this value depends on the choice of $\tau$, $V$ and $\Gzero$ are discussed in Sec.~\ref{sec:results_robustness}.
This also explains the previous paragraph quantitatively: raising the reference timescale by five orders of magnitude raises the required reduced barrier only by $\ln(10^5)\simeq12$, which accounts for the shift $\simeq104\to107\MeVfm$ in $\sigcrit$ quoted above.

Table~\ref{tab:outcomes} collects representative outcomes for both parametrizations, from $1.4\Msol$ up to the maximum-mass configuration $\MmaxPNS(\tzero)=1.78\Msol$. Because baryon number is conserved during conversion, the QS remnant is fixed by baryon matching on the $M$--$\MB$ curves of Fig.~\ref{fig:seq}(b), and the released energy is $\Econv=\MPNS-\MQS$. The table also records the epoch at which the barrier is first overcome: as $\sigma$ increases, conversion is delayed from $\tzero$ to $\tTmax$ and eventually suppressed. Three outcomes arise. Low- and intermediate-mass PNSs that fail to nucleate cool into NSs of equal baryonic mass. The maximum-mass configuration, by contrast, has a baryonic mass exceeding the maximum that can be supported at $\tTmax$; if it fails to nucleate at $\tzero$ it can neither convert nor reach a stable later configuration, and collapses to a black hole before $\tTmax$. We find $\Econv\simeq(4.4$--$5.4)\times10^{53}$~erg for Set~A and $(6.5$--$8.0)\times10^{53}$~erg for the more strongly bound Set~B, potentially observable as a delayed neutrino/gravitational-wave transient or a GRB (Sec.~\ref{sec:conversion_nucleation_remnant}).

The table settles the outcome for two particular choices of the SQM parameters. Whether \emph{any} choice leaves room for an ordinary $1.4\Msol$ NS to survive is the question of the next section.

\begin{table*}
  \centering
  \small
  \setlength{\tabcolsep}{3.5pt}
  \begin{adjustbox}{max width=\textwidth}
  \begin{tabular}{cc|cccc|ccc|ccc}
    \hline\hline
    \multicolumn{2}{c|}{} &
    \multicolumn{4}{c|}{PNS at $\tzero$} &
    \multicolumn{3}{c|}{PNS at $\tTmax$} &
    \multicolumn{3}{c}{Remnant} \\
    Set & $\sigma$ &
    $\MPNS$ & $\MB^{\rm PNS}$ & $\nBcH$ & $Y_{S,c}^{H}$ &
    $\MPNS$ & $\nBcH$ & $Y_{S,c}^{H}$ &
    type & $M$ & $\Econv$ \\
    & $[\MeVfm]$ &
    [M$_\odot$] & [M$_\odot$] & [$\nsat$] & &
    [M$_\odot$] & [$\nsat$] & &
    & [M$_\odot$] & [$\times10^{53}$ erg] \\
    \hline
    \multirow{9}{*}{A}
    & 50 & 1.40 & 1.48 & 3.00 & 0.01 & -- & -- & -- & QS & 1.15 & 4.51 \\
    &  & 1.60 & 1.72 & 3.85 & 0.03 & -- & -- & -- & QS & 1.32 & 5.05 \\
    &  & 1.78 & 1.95 & 6.63 & 0.21 & -- & -- & -- & QS & 1.47 & 5.45 \\
    & 100 & 1.40 & 1.48 & 3.00 & 0.01 & 1.39 & 3.21 & 0.03 & QS & 1.15 & 4.41 \\
    &  & 1.60 & 1.72 & 3.85 & 0.03 & 1.59 & 4.31 & 0.07 & QS & 1.32 & 4.92 \\
    &  & 1.78 & 1.95 & 6.63 & 0.21 & -- & -- & -- & QS & 1.47 & 5.45 \\
    & 150 & 1.40 & 1.48 & 3.00 & 0.01 & 1.39 & 3.21 & 0.03 & NS & 1.34 & -- \\
    &  & 1.60 & 1.72 & 3.85 & 0.03 & 1.59 & 4.31 & 0.07 & NS & 1.53 & -- \\
    &  & 1.78 & 1.95 & 6.63 & 0.21 & -- & -- & -- & BH & -- & -- \\
    \hline
    \multirow{9}{*}{B}
    & 150 & 1.40 & 1.48 & 3.00 & 0.01 & -- & -- & -- & QS & 1.03 & 6.54 \\
    &  & 1.60 & 1.72 & 3.85 & 0.03 & -- & -- & -- & QS & 1.19 & 7.35 \\
    &  & 1.78 & 1.95 & 6.63 & 0.21 & -- & -- & -- & QS & 1.33 & 8.00 \\
    & 200 & 1.40 & 1.48 & 3.00 & 0.01 & 1.39 & 3.21 & 0.03 & NS & 1.34 & -- \\
    &  & 1.60 & 1.72 & 3.85 & 0.03 & 1.59 & 4.31 & 0.07 & QS & 1.19 & 7.22 \\
    &  & 1.78 & 1.95 & 6.63 & 0.21 & -- & -- & -- & QS & 1.33 & 8.00 \\
    & 250 & 1.40 & 1.48 & 3.00 & 0.01 & 1.39 & 3.21 & 0.03 & NS & 1.34 & -- \\
    &  & 1.60 & 1.72 & 3.85 & 0.03 & 1.59 & 4.31 & 0.07 & NS & 1.53 & -- \\
    &  & 1.78 & 1.95 & 6.63 & 0.21 & -- & -- & -- & BH & -- & -- \\
    \hline\hline
  \end{tabular}
  \end{adjustbox}
  \caption{Representative outcomes of PNS evolution for Set~A ($\Bq=145\MeV$, $\Dzero=80\MeV$, $\alphas=\tfrac{\pi}{2}\times0.1$) and Set~B ($\Bq=158\MeV$, $\Dzero=157\MeV$, $\alphas=\tfrac{\pi}{2}\times0.08$), in the unpCFL scheme, as a function of the surface tension $\sigma$. The central baryon density $\nBcH$ and strangeness fraction $Y_{S,c}^H$ of the PNS are given at both the $\tzero$ and $\tTmax$ snapshots; the $\tTmax$ block is left blank (``--'') when the star has already nucleated at $\tzero$ and never reaches that epoch. The outcome is QS (conversion to a QS of equal baryonic mass), NS (cooling onto the cold NS branch of equal baryonic mass), or BH (gravitational collapse). The conversion energy is $\Econv=\MPNS-\MQS$.}
  \label{tab:outcomes}
\end{table*}

\subsection{The NS--QS coexistence window in the SQM parameter space}
\label{sec:results_paramspace}

\begin{figure*}
    \centering
    \includegraphics[width=1\textwidth]{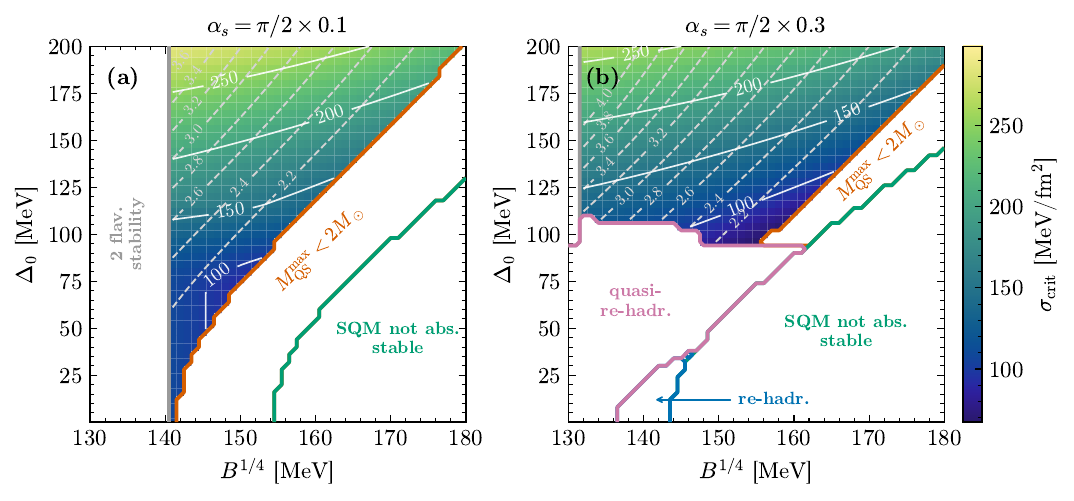}
    \caption{Critical surface tension $\sigcrit$ (color map) in the $(\Bq,\Dzero)$ plane for two values of the strong coupling, $\alphas=\tfrac{\pi}{2}\times0.1$ [panel~(a)] and $\tfrac{\pi}{2}\times0.3$ [panel~(b)], in the unpCFL scheme. $\sigcrit$ is the surface tension above which a PNS with the baryonic mass of a cold $1.4\Msol$ NS fails to nucleate at $\tTmax$ and survives as a NS. White dashed contours are isolines of $\MmaxQS/\Msol$ while white solid lines are $\sigcrit$ isolines. The accepted island is bounded by the onset of two-flavor stability (gray), the loss of three-flavor SQM absolute stability (green), and $\MmaxQS=2\Msol$ (orange); for panel~(b), additional re-hadronization (blue) and quasi-re-hadronization (pink) boundaries appear at low $\Dzero$.}
    \label{fig:sigcrit}
\end{figure*}

We now ask whether a region of the SQM parameter space exists in which some hadronic PNSs do not reach the nucleation conditions that would trigger their conversion into absolutely stable QSs, and cool down into metastable NSs. This is a necessary condition for the coexistence of the two families of compact stars. 
We scan the space of SQM parameters $(\alphas,B,\Dzero)$ on a regular grid, $\Bq\in[130,180]\MeV$, $\Dzero\in[0,200]\MeV$  and $\alphas=\tfrac{\pi}{2}\times(0.1,0.2,0.3)$ at fixed $\ms=100\MeV$, and retain only the parametrizations that satisfy some physical requirements, each of which excludes part of the plane and so traces one of the boundaries of the accepted island in Fig.~\ref{fig:sigcrit}: 
\begin{itemize}
\item \emph{Absolute stability of SQM} (the BW hypothesis), a prerequisite for the two-families scenario and for the existence of QSs: the green boundary marks the loss of SQM absolute stability; to its right, the $E/A$ of SQM exceeds that of $^{56}$Fe. 
\item \emph{Non-absolute stability of two-flavor quark matter}: the gray band marks the complementary two-flavor stability line, to the left of which non-strange quark matter would already be stable and the SQM hypothesis would lose its meaning. 
\item \emph{No re-hadronization}, i.e.\ once cold bulk SQM becomes more stable than cold hadronic matter, it stays so at every higher pressure. Operationally we build $\Delta P(\mu_B)=P_{\rm CFL}-\PH$ at $T=0$, between the CFL quark EOS and the cold, neutrinoless, $\beta$-equilibrated hadronic one, and require $\Delta P>0$ at every $\mu_B$ above the first crossing (which, if present, is at $P<0$ in the BW hypothesis). 
\item \emph{No ``quasi-re-hadronization''}: the same $\Delta P(\mu_B)$, as well as its counterpart for the hot, lepton-rich matter of PNSs, is further required to be monotonically increasing, so that SQM becomes progressively more favored as the pressure grows.
\item \emph{Massive-pulsar support}, $\MmaxQS>2\Msol$: the orange boundary marks $\MmaxQS=2\Msol$, beyond which the QS branch is too soft to support the observed massive pulsars.
\end{itemize}
The two conditions on re-hadronization are not strong theoretical requirements, but physically reasonable regularity conditions that exclude EOSs in which the phase competition reverses at high $P$. They generate respectively the blue and pink boundaries that appear at low $\Dzero$ for the larger coupling [panel~(b)]. 
The massive-pulsar requirement is a minimal astrophysical constraint; it could be superseded by a full Bayesian inference of the QS EOS~\cite{Passarella:2026mxy}. 

For every parametrization that passes these cuts we compute the critical surface tension $\sigcrit$ that just prevents nucleation in a reference PNS. In Sec.~\ref{sec:results_nucleationPNS} the critical surface tension was defined as a function of the stellar mass, $\sigcrit(\MPNS)$, at the $\tTmax$ snapshot. Throughout this section we evaluate $\sigcrit$ as the surface tension above which the nucleation condition ($\tau\leq \taudyn =1\,$ms) is no longer met at the $\tTmax$ snapshot, in a PNS whose baryonic mass equals that of a cold NS of gravitational mass $1.4\Msol$ (i.e.\ $\MPNS\simeq1.46\Msol$ at that epoch, see the corresponding open dot in Fig.~\ref{fig:Tnuc}); below $\sigcrit$ that star nucleates and no $1.4\Msol$ NS can exist. A weaker choice would set the reference mass at $\simeq1.17\Msol$, the lightest NS produced in core-collapse supernovae~\cite{Suwa:2018uni,Muller:2024aod}; we adopt the stronger $1.4\Msol$ threshold because one of the strengths of the two-families scenario is precisely its account of the small radii inferred for ordinary $\sim1.4\Msol$ NSs. However, the effect of relaxing this choice is quantified at the end of the section.

Figure~\ref{fig:sigcrit} shows $\sigcrit$ (color map) in the $(\Bq,\Dzero)$ plane, for two values of the strong coupling ($\alphas=\tfrac{\pi}{2}\times0.1$ and $\tfrac{\pi}{2}\times0.3$), together with the exclusion boundaries introduced above and white dashed isolines of $\MmaxQS/\Msol$. Across the accepted island $\sigcrit$ ranges from $\simeq67$ to $\simeq287\MeVfm$; the upper end is set by the edge of the scanned range, $\Dzero=200\MeV$.

The map makes the interplay of the microphysical parameters explicit. At fixed $\alphas$, a larger gap $\Dzero$ has two effects: it increases SQM stability, raising $\MmaxQS$, favoring the $2\Msol$ constraint, but it also shrinks the pairing radius $\Rgap$ and thereby eases nucleation, so that a larger $\sigcrit$ is needed to preserve NS survival; $\sigcrit$ therefore increases toward the top of each panel. Increasing $\alphas$ shifts the stability boundaries and displaces the accepted island toward higher $\Dzero$: for $\alphas=\tfrac{\pi}{2}\times0.1$ the island reaches down to $\Dzero\simeq0$, whereas for $\alphas=\tfrac{\pi}{2}\times0.3$ it requires $\Dzero\gtrsim96\MeV$.
The bag constant acts in the opposite sense. At fixed $\Dzero$, a larger $\Bq$ makes SQM less bound: $\MmaxQS$ falls and nucleation is harder, so $\sigcrit$ decreases from left to right across each panel. 
The two parameters are not independent, however, because it is their combination that fixes $\MmaxQS$, which is read off Fig.~\ref{fig:sigcrit} by following a white dashed contour rather than a row or a column: moving along a line of constant $\MmaxQS$ toward larger $\Dzero$ (and hence, to keep the mass fixed, larger $\Bq$) makes $\sigcrit$ rise. Along $\MmaxQS=2.2\Msol$ at $\alphas=\tfrac{\pi}{2}\times0.1$, $\sigcrit$ grows from $\simeq105\MeVfm$ at $(\Bq,\Dzero)=(145,80)\MeV$, i.e.\ Set~A, to $\simeq 238\MeVfm$ at $(173,200)\MeV$; the $2.4$ and $2.6\Msol$ contours behave the same way, spanning $\simeq 118$--$249$ and $\simeq 150$--$258\MeVfm$. Two SQM parametrizations that are indistinguishable by the maximum mass of the QS branch can therefore demand surface tensions differing by more than a factor of two before a $1.4\Msol$ NS is allowed to survive. 

The map of Fig.~\ref{fig:sigcrit} is computed in the unpCFL scheme; using instead the pure-CFL prescription changes $\sigcrit$ only where the critical droplet is not already CFL. 
At high $\Dzero$ the gap radius $\Rgap$ is small, the critical droplet is already CFL, and the two prescriptions coincide exactly: the island is dominated by regime~(iii), $\Rs=\Rscfl$. 
The difference is therefore confined to a band in $\Dzero$, comprising the gap-locked regime~(ii), $\Rs=\Rgap$, and part of the unpaired regime~(i), $\Rs=\Rsunp$. There the unpCFL barrier is higher than the pure-CFL one, so that a \emph{lower} surface tension already suffices to suppress nucleation, $\sigcrit^{\rm unpCFL}<\sigcrit^{\rm CFL}$. For $\alphas=\tfrac{\pi}{2}\times0.1$ the two prescriptions differ in cells confined to $\Dzero\simeq 50$--$116\MeV$ and $\Bq\simeq141$--$157\MeV$; over those cells the median shift is $-13\MeVfm$ and the largest is $-21\MeVfm$. 
For $\alphas=\tfrac{\pi}{2}\times0.3$ only two cells are affected, at the lower edge of the island ($\Dzero=108\MeV$), by $-1\MeVfm$.  The unpCFL treatment is thus essential for the low- and intermediate-coupling, intermediate-gap corner, where it lowers the minimum of each slice, from $\simeq99$ to $\simeq93\MeVfm$ for $\alphas=\tfrac{\pi}{2}\times0.1$ and from $\simeq89$ to $\simeq75\MeVfm$ for $\tfrac{\pi}{2}\times0.2$, and nearly immaterial for large $\alphas$. The minimum of the whole island, $\sigcrit\simeq67\MeVfm$, lies at $(\Bq,\Dzero)=(155,96)\MeV$ and $\alphas=\tfrac{\pi}{2}\times0.3$: there the critical droplet is already CFL [regime~(iii)], so the lower bound on $\sigma$ does not depend on the finite-size treatment of pairing.
Below that $\Dzero$ band, the pairing gap melts thermally and goes to zero once $\TcCFL<\Tcen$, i.e.\ for $\Dzero\lesssim46\MeV$ at the central temperature $\Tcen\simeq33\MeV$ of the reference star. There $\Rgap\to\infty$ and the unpaired phase is the stable SQM phase, thus the unpCFL droplet is simply in the unpaired regime~(i), $\Rs=\Rsunp$.

The inferred $\sigcrit\gtrsim 67 \MeVfm$ is large compared with the lowest quark-matter surface tensions considered in the literature ($\sim10$--$30\MeVfm$), but it lies within the broad range spanned by microscopic estimates, several of which favor values of a few tens to a few hundred $\MeVfm$ (see Sec.~\ref{sec:conversion_nucleation_energybarrier}).

Finally, the massive-pulsar boundary can be sharpened if a specific massive object is interpreted as a QS. Requiring $\MmaxQS>2.18\Msol$, the $-1\sigma$ mass of PSR~J0952$-$0607~\cite{Romani:2022jhd}, already raises the minimum $\sigcrit$ to $\simeq81\MeVfm$. Requiring $\MmaxQS>2.35\Msol$, the central value for the mass of the same pulsar, restricts the island to $\Dzero\gtrsim 84\MeV$ and raises the minimum $\sigcrit$ to $\simeq 94\MeVfm$. Interpreting the secondary of GW190814 ($\simeq2.6\Msol$) as a QS is far more demanding, requiring $\Dzero\gtrsim92\MeV$, $\Bq\lesssim164\MeV$, and $\sigcrit\gtrsim 110\MeVfm$. A heavier maximum mass thus forces both a larger diquark gap and a larger surface tension, tightening the coexistence window.

We finally want to quantify the dependence on the reference mass. Adopting the weaker threshold $1.17\Msol$ instead of $1.4\Msol$  lowers $\sigcrit$ across the island by a median $-17\MeVfm$ (up to $-31\MeVfm$), because a lighter reference star has a lower central density and nucleates less readily. Requiring instead that a $1.6\Msol$ NS survive raises $\sigcrit$ by a median $+25\MeVfm$ (up to $+55\MeVfm$). In either case the shape of the accepted island and the qualitative parameter trends are unchanged. The minimum of the island, however, barely moves: $\sigcrit\simeq66$, $67$ and $67\MeVfm$ for $1.17$, $1.4$ and $1.6\Msol$.

Since the hadronic EOS has been kept fixed, the bound on $\sigma$ is not universal: it refers to the specific combination of hadronic and quark models adopted here.
The hadronic EOS enters the calculation in three places. First, it fixes the central conditions of the reference PNS: a softer hadronic EOS produces a more compact star, with a denser but, at fixed entropy per baryon, cooler center. Second, it sets the hadronic side of the driving force $P_{\Qs}-\PH$: an earlier onset of hyperons or $\Delta$ resonances raises the hadronic pressure at given $\mu_B$, thus weakening the driving force at fixed SQM parameters. Third, through the (quasi-)re-hadronization conditions, it shapes the lower edge of the accepted island at large $\alphas$, where the minimum of $\sigcrit$ lies: a hadronic EOS with a larger baryon density at given $\mu_B$ makes $\Delta P(\mu_B)$ non-monotonic over a wider range of SQM parameters, removing the low-$\Dzero$ corner of the island. Since these effects act in different directions, their net impact on the bound can only be assessed through a consistent exploration of the hadronic parameter space, which we leave for future work.

The result of this section can then be stated compactly: within the $\alpha$Bag description of SQM and for the SFHo-H$\Delta$ hadronic EOS, the coexistence of NSs and QSs requires a hadron--quark surface tension of at least $\sigma\simeq 67\MeVfm$ in the accepted parameter space. This is a lower bound, not a preferred value: any $\sigma$ below it converts every PNS heavy enough to become a canonical $\sim1.4\Msol$ NS, thereby disfavoring the two-families scenario. 
Some caveats, and a brief discussion of how relaxing the assumptions made in this work would affect this bound, are deferred to the conclusions.

\subsection{Robustness of the nucleation criterion}
\label{sec:results_robustness}

The nucleation conditions, and hence $\sigcrit$, rest on three inputs that are conventions or approximations rather than properties of the matter: the reference timescale $\tau$, the nucleation volume $V$ and the rate prefactor $\Gzero$. Since Eqs.~\eqref{eq:nucreate} and~\eqref{eq:taunuc} give $\Ws/T=\ln(\tau V\Gzero)$, their effects can be discussed together. Splitting the logarithm around the reference choice $\tau=\taudyn=1$~ms and $V=V^{\rm PNS}_{\rm center}$,
\begin{equation}
  \begin{split}
 \frac{\Ws}{T}
  &=\ln\!\left[1~{\rm ms}\;V^{\rm PNS}_{\rm center}\,\Gzero\right]
   +\ln\!\left[\frac{\tau}{1~{\rm ms}}\,\frac{V}{V^{\rm PNS}_{\rm center}}\right]\\
  &\simeq167+\ln\!\left[\frac{\tau}{1~{\rm ms}}\,\frac{V}{V^{\rm PNS}_{\rm center}}\right].
  \end{split}
  \label{eq:WTrescale}
\end{equation}
The constant $167$ is the sum of $\ln(1~{\rm ms}\;V^{\rm PNS}_{\rm center}/{\rm s\,fm^3})\simeq112$, fixed by the reference choice, and $\ln[\Gzero/({\rm fm^{-3}\,s^{-1}})]\simeq54$--$56$. This split depends on the units while the scatter comes from the prefactor, the only term that depends on the matter conditions. Over the accepted island of Fig.~\ref{fig:sigcrit}, at the reference $\tau$ and $V$, the constant has a median of $167.4$, with $68\%$ ($95\%$) of the cells between $166.9$ and $168.4$ ($166.6$ and $170.9$).

Raising $\tau$ from $1$~ms to $100$~s, an upper bound on the duration of the hot PNS phase, raises the threshold $\Ws/T$ by $\simeq 12$ and $\sigcrit$ by $\simeq 3$--$5\MeVfm$.
Integrating $\Gamma$ over the proper volume of the representative stars, with $\nBH(r)$ and $m(r)$ from the TOV equations and $T(r)$ from the isentrope, the rate is maximal at the center, but $\Ws/T$ actually grows slowly enough outward that the equivalent uniform-rate region has a radius of $0.6$--$1.7$~km, i.e.\ a volume $\sim 10^{2}$--$10^{4}$ times $V^{\rm PNS}_{\rm center}$. The threshold rises to $\Ws/T\simeq 173$--$177$ and $\sigcrit$ by $\simeq 2\MeVfm$.
Replacing the prefactor with the dimensional estimate $\Gzero\to T^4$, often adopted in the literature, lowers $\ln\Gzero$ by $7.5$--$11$ at the central temperatures of the representative stars ($T\simeq23$--$34\MeV$), hence the threshold to $\Ws/T\simeq158$, and $\sigcrit$ by $\simeq2$--$4\MeVfm$.

All these shifts amount to a few $\MeVfm$, thus small compared with the physical uncertainties quantified above: the finite-size treatment of pairing moves $\sigcrit$ by up to $21\MeVfm$, the choice of the reference mass by a median $-17$ or $+25\MeVfm$ (Sec.~\ref{sec:results_paramspace}), while $\sigcrit$ itself spans $67$--$287\MeVfm$ across the accepted island.

\section{Summary and Conclusions}
\label{sec:conclusions}

The aim of this work was to establish whether a region of the SQM parameter space exists in which NSs and QSs can coexist, as required by the two-families scenario, built on the BW hypothesis. The issue that motivates the question is that, if SQM is absolutely stable, a hadronic NS is a metastable stellar object, and one has to explain how it can survive for astronomical times without converting into a QS. Since every compact star begins its life as a hot, lepton-rich PNS, and since the conditions reached at that stage are more favorable to deconfinement than those of a cold star, the existence of NSs requires that at least part of the PNS population complete its cooling and deleptonization without nucleating SQM.

We have therefore computed the thermal nucleation of a critical SQM droplet under the conditions of two representative snapshots of PNS evolution, taken at the time of the CCSN explosion and at the epoch of maximum central temperature. Two ingredients distinguish the present calculation from the standard treatment. First, the composition of the droplet is not frozen to that of the surrounding average bulk hadronic matter: it is fixed by a saddle-point condition on the work of formation, which is equivalent to strong chemical equilibrium across the interface and which can be realized either by a prior local fluctuation of the hadronic composition or by the exchange of flavor across the droplet surface. Second, color superconductivity is treated at finite droplet size: because diquark pairs cannot form in a region smaller than their coherence length $\Rgap=1/\Delta$, the droplet is in the unpaired phase below $\Rgap$ and in the CFL phase above it.

Since the nucleation time varies very fast with the baryon density and temperature, it is meaningful to speak of \emph{nucleation conditions}, defined as the thermodynamic states at which $\tau$ equals the dynamical timescale $\taudyn=1$~ms, and the precise value adopted for the reference timescale turns out to be immaterial: varying it over five orders of magnitude, up to $100$~s, an upper limit for the duration of the hot PNS phase, shifts the threshold surface tension negligibly with respect to the other uncertainties of the problem. The same holds for the nucleation volume: integrating the rate over the whole star rather than over a central sphere of radius $100$~m raises $\sigcrit$ by only $\simeq2\MeVfm$ (Sec.~\ref{sec:results_robustness}). 
Comparing the nucleation conditions with the ones actually reached by PNSs identifies the epoch of maximum central temperature $\tTmax$ as the bottleneck for the fate of the star: a PNS that has not nucleated by then will cool into a NS. This is what allows the \emph{critical surface tension} $\sigcrit$ to be defined as the value of $\sigma$ above which the nucleation condition is no longer met at $\tTmax$ inside a PNS whose baryonic mass equals that of a cold $1.4\Msol$ NS: for $\sigma<\sigcrit$ that PNS nucleates,  and no $1.4\Msol$ NS can exist.

Our main results can be summarized as follows.

For the combination of hadronic and quark EOSs adopted here, the coexistence of the two families requires a hadron--quark surface tension of at least $\sigma\simeq67\MeVfm$: below that value, every PNS heavy enough to become an ordinary $1.4\Msol$ NS converts. This bound depends only marginally on the choice of the reference mass ($\simeq66$--$67\MeVfm$ between $1.17$ and $1.6\Msol$). This minimum is attained by the softest SQM parametrizations still compatible with $2\Msol$ pulsars, thus it tightens whenever a heavier object is identified with a QS: to $\simeq$ $81$, $94$ and $110\MeVfm$ if the QS branch is required to reach $2.18$, $2.35$ and $2.6\Msol$, the first two being the $-1\sigma$ and the central mass of PSR~J0952$-$0607~\cite{Romani:2022jhd} and the third being the mass of the secondary component of GW190814~\cite{LIGOScientific:2020zkf}.
These are lower bounds and not preferred values; they are large compared with the smallest surface tensions considered in the literature, but well inside the range spanned by microscopic estimates. We stress that $\sigcrit$ is not fixed by the maximum mass of the QS branch alone: along a line of constant $\MmaxQS$ it varies by more than a factor of two, so that two parametrizations indistinguishable by their maximum mass can demand very different surface tensions.

The finite-size treatment of pairing has a qualitative impact under some conditions. Depending on density and temperature, the critical droplet is fully unpaired, gap-locked at $\Rs=\Rgap$, or fully CFL, and neglecting the geometric suppression of pairing changes the nucleation time by many orders of magnitude and the way the barrier responds to the surface tension. Its effect on $\sigcrit$ is confined to an intermediate gap band of the parameter space, with an impact of at most $\simeq21$ $\MeVfm$.

Two further numbers are worth carrying over. At the nucleation conditions the reduced barrier is nearly universal, $\Ws/T\simeq167+\ln[(\tau/1~{\rm ms})(V/V^{\rm PNS}_{\rm center})]$ [Eq.~\eqref{eq:WTrescale}] to within a few units, so that a local thermodynamic state, for instance a simulation cell of volume $V$ over a time step $\tau$, can be approximately tested for deconfinement without computing the rate. Moreover, when conversion does occur it releases several $\times10^{53}$~erg, depending on the SQM parametrization and on the mass of the converting star. This energy reservoir could power a delayed transient after a CCSN.

Several limitations should be kept in mind, and each of them defines a natural direction for future work.

First, we have treated $\sigma$ as a constant free parameter and neglected the curvature term, whereas both should be computed consistently from the underlying models and both depend in general on density and temperature. A fully consistent calculation would require a single potential describing the hadronic and the quark phase, while we are using two separate models; a multiple-reflection-expansion estimate would be a first step, although, as discussed in Sec.~\ref{sec:conversion_nucleation_energybarrier}, different microscopic calculations still give widely different values. 

The treatment of pairing in small droplets is deliberately minimal: a sharp switch at $\Rgap=1/\Delta$ stands in for what is really a condensate that turns on gradually with the size of the system. A more complete analysis of diquark pairing in finite systems might modify the results: if the true coherence length were larger than assumed, the critical droplet would remain unpaired over a wider range of conditions, the barrier would be higher and a smaller surface tension would already suffice to prevent nucleation. The extreme case is a first droplet that never pairs at all: for the parametrization of Fig.~\ref{fig:sigcrit_mass} this lowers $\sigcrit$ at $1.4\Msol$ from $\simeq 104$ to $\simeq99\MeVfm$.

Another, more structural caveat concerns which droplet configurations are actually reachable. The saddle-point treatment of the composition assumes that every configuration compatible with the strong-interaction conservation laws (namely, those that do not require weak interactions) can be realized, either through a prior fluctuation of the hadronic composition or through flavor exchange across the interface while the droplet exists. In a more realistic treatment of such a complicated dynamical system, one should also consider a phenomenon not taken into account within a purely thermodynamical approach: rearranging a conserved charge over the size of the droplet takes time, and configurations that would require transporting it further than the available diffusion length should be excluded from the accessible ensemble. Restricting the accessible configurations can only raise the barrier, so in this respect the present results are conservative and may overestimate the ease of nucleation. A preliminary framework for taking care of this effect has been presented in~\cite{Guerrini:2025ojt}, where the restriction is found to matter most precisely when the hadronic phase carries little strangeness; a quantitative treatment requires the use of transport properties in hadronic and quark matter that are beyond the scope of this work.

We have also relied on a single hadronic EOS parametrization and on one model for the quark phase, so the numerical value of the bound on $\sigma$ should not be regarded as universal. The hadronic EOS, although compatible with the two-families posterior of~\cite{Passarella:2026mxy}, has been kept fixed: our results therefore establish that, for at least one hadronic EOS compatible with that analysis, there exist SQM parametrizations and surface tensions for which a canonical $1.4\Msol$ NS survives the PNS stage, as required by the two-families scenario. Within the $\alpha$Bag description the pairing gap is a constant parameter, whereas in a model such as NJL it is a quantity to be computed and is density dependent, which would make $\Rgap$ density dependent as well and would modify the extent of the gap-locked regime. We have moreover used the same bag constant for the unpaired and the CFL phase, which is the natural choice but not a necessary one; allowing the two values of the bag constant to differ would open a further direction in parameter space. Testing the framework against other hadronic and quark EOSs is therefore an obvious next step.

The thermodynamic conditions of the PNS have been taken from simulations performed with a different hadronic EOS, and summarized by two representative snapshots. The same uniform $\SH/N_B$ and $\YLH$ enter the whole-star integral of the rate of Sec.~\ref{sec:results_robustness}; in a real PNS both vary with radius, which changes how fast $\nBH$ and $T$ fall off away from the center and hence the size of the nucleating region, although by Eq.~\eqref{eq:WTrescale} this affects $\sigcrit$ only logarithmically. Dedicated simulations with the EOS and the parametrizations used here would replace this approximation with an actual evolutionary track.

Finally, we have imposed only minimal astrophysical requirements, namely that the QS branch support $2\Msol$ pulsars. A natural development is to embed the present nucleation criterion in a full Bayesian inference of the joint hadronic--SQM parameter space~\cite{Passarella:2026mxy}, replacing the qualitative assumption adopted there (namely, that any star reaching a central strangeness density $n_S\geq\nsat$ nucleates and cannot persist as a metastable configuration~\cite{DePietri:2019khb,Bombaci:2020vgw,Becerra:2025ryy}) by the conditions derived in this work. Such an exhaustive exploration is beyond the aim of the present paper and it is an interesting direction for future research.

The same framework applies well beyond the PNS stage. Nucleation could also be triggered during the rapid accretion that follows core bounce, in BCSMs, or in accreting binaries with a NS companion. Unlike PNSs and mergers, accreting NSs are cold, $T\lesssim 0.1\MeV$, so that quantum tunneling rather than thermal activation would dominate the decay of the metastable phase. Beyond the BW hypothesis, the same finite-size treatment can be used for the formation of hybrid stars, and, with the appropriate changes in system size and particle content, for heavy-ion collisions.

We close with a caveat of a different kind. We have assumed throughout that the transition proceeds by nucleation, that is, that the system crosses a genuine barrier from a metastable state. If instead an external perturbation drove the matter into the mechanically unstable region faster than a critical droplet can form, the transition would start by spinodal decomposition, with no barrier to overcome and no critical size: any fluctuation would grow immediately. Our approach is blind to this possibility, since a proper treatment requires a single potential describing both phases~\cite{Bessa:2008nw}. In a PNS, however, the thermodynamic conditions evolve slowly enough that the star should remain in the metastable region long enough for critical droplets to form before the unstable region is reached, thus we expect nucleation to be the relevant channel.

\section*{Data Availability}
The source code, the Python notebook employed to generate the reported results, and additional supplementary materials are available at \href{https://github.com/guerrinimirco/metastability-nucleation.git}{github.com/guerrinimirco/metastability-nucleation.git}.

\bibliography{apssamp}

\appendix

\section{Electric charge neutrality and the Coulomb term}
\label{app:screening}

The critical droplet may carry a net electric charge, and how that charge is treated changes both $\Rs$ and $\Ws$. Whether neutrality is effectively local or only global is set by the comparison between the droplet size and the Debye screening length $\lambda_D$: for $\Rs\gg\lambda_D$ the mobile charges screen the interface efficiently and each phase relaxes close to local neutrality, whereas for $\Rs\lesssim\lambda_D$ a sizable charge separation survives and only global neutrality holds. At the conditions considered below neither limit applies cleanly, $\Rs/\lambda_D\simeq0.5$--$0.6$ with $\lambda_D\simeq7.3\fm$, so it is worth quantifying the spread between them. We compare five prescriptions:
\begin{itemize}
\item \emph{LCN}: local charge neutrality is imposed inside the droplet, $Y_C^{\Qs}-Y_e^{\Qs}=0$, so that the Coulomb energy vanishes identically;
\item \emph{GCN}: the droplet is allowed to be charged, with $\mu_e^{\Qs}=\mu_e^{H}$, and the Coulomb energy is neglected;
\item \emph{GCN\,+\,Coulomb}: the unscreened Coulomb energy of a uniformly charged sphere, $\Wcoul\propto\dnC^2\Rs^5$, is added \emph{after} the composition has been fixed;
\item \emph{minimization}: the Coulomb energy is included \emph{inside} the minimization of $\Ws$, which shifts $\mu_e^{\Qs}$ by a term proportional to $\dnC\Rs^2$ [Eq.~\eqref{eq:muemueminim}] and thereby penalizes charged droplets. This is the scheme adopted throughout the paper;
\item \emph{screening}: the electrons are allowed to redistribute, the linearized Poisson equation is solved, and the Coulomb energy acquires the Debye form of~\cite{Guerrini:2025ojt,Iida:1997ay,Iida:1998pi}.
\end{itemize}

\begin{figure*}
    \centering
    \includegraphics[width=1\textwidth]{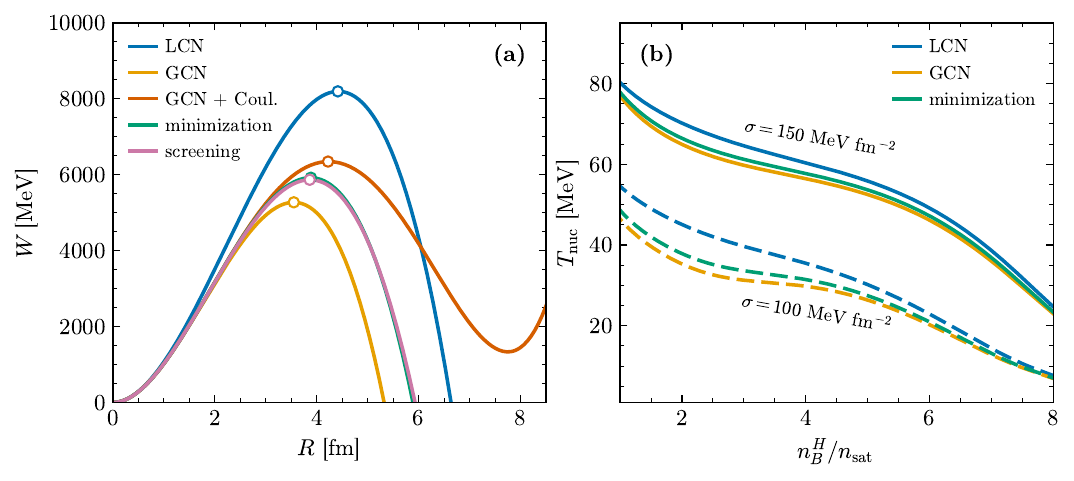}
    \caption{(a) Work of formation $W(R)$ of an unpaired droplet under the five prescriptions discussed in the text, at $\nBH=3.1\,\nsat$, $T=30\MeV$, $\YLH=0.25$, $\sigma=100\MeVfm$ and Set~A. Circles mark the critical point of each curve. 
    (b) Nucleation temperature $\Tnuc$ versus $\nBH/\nsat$ for LCN, GCN and minimization in the purely unpaired case, for Set~A at $\YLH=0.25$ and two surface tensions, $\sigma=100$ and $150\MeVfm$ (color: prescription; line style: $\sigma$).}
    \label{fig:app_charge}
\end{figure*}

In Fig.~\ref{fig:app_charge}(a) we show the energy barrier as a function of the radius for the purely unpaired case with the different approaches presented in this appendix. The blue line assumes local charge neutrality, the yellow line assumes global charge neutrality neglecting the role of Coulomb energy, and the orange line simply adds the Coulomb energy to the global charge neutrality case. Finally, the green and pink lines are the more sophisticated approaches. The former assumes that the particle densities are constant inside and outside the droplet (but different between inside and outside) and their values are optimized to minimize $W$ (including the Coulomb energy). In the latter the electron density is continuous and not constant since electrons act to screen the net electric charge. 

First, note that the GCN\,+\,Coulomb case has no critical radius, since the $\propto R^5$ term makes the second minimum lie at higher energy than the $R=0$ one; the maximum therefore does not correspond to a critical droplet.
In the green and pink lines, the electric charges are free to reallocate, which reduces the Coulomb energy at the price of a higher bulk term. The result is that the qualitative shape of the barrier is similar to the LCN and GCN (without Coulomb interaction) ones, but lies between these two cases.

In Fig.~\ref{fig:app_charge}(b) we show the nucleation temperature as a function of $n_B^H$ for the purely unpaired case at $\sigma=100\MeVfm$ and $150\MeVfm$.
The yellow and blue lines, as in panel~(a), represent the GCN and LCN cases, while the green lines correspond to the approach used in the paper. The different approaches have a relevant impact at high $T$, while they converge to the same result at low $T$ and high $n_B^H$. Again, the minimization approach gives results between the GCN and LCN ones.

\section{Composition of the critical droplet}
\label{app:composition}

The second methodological choice worth isolating is the composition of the droplet. In the \emph{frozen-flavor} prescription, used in much of the literature, the droplet inherits the average flavor content of the surrounding bulk hadronic matter, $Y_C^{\Qs}=Y_C^{H}$ and $Y_S^{\Qs}=Y_S^{H}$: the exchange of strong charges across the interface is assumed to be ineffective on the lifetime of the fluctuation, and the hadronic phase is assumed to be compositionally homogeneous. 
In the \emph{saddle-point} prescription adopted here, both assumptions are relaxed and the composition is the one that minimizes the work of formation, which is equivalent to strong chemical equilibrium across the interface.

Physically, the configuration that minimizes the barrier can be reached with two different mechanisms or a combination of them. The hadronic phase can itself fluctuate before the droplet appears, producing a subsystem whose local flavor content differs from the bulk average; nucleation then takes place inside that subsystem, and the total work of formation is the sum of the cost of creating the fluctuation and the cost of creating the droplet within it. Alternatively, the droplet can exchange net flavor with its surroundings across the interface while it exists. Either way, the accessible configurations are those reachable by the strong interaction alone, and a complete calculation of the rate would integrate over all of them. Because the rate is exponential in $-\Ws/T$, that integral is dominated by the single configuration of least work, and it is that configuration, the saddle point, that we use. 

Thus, the two prescriptions are two limits of the same question: which droplet configurations are actually reachable in a system with a finite lifetime. The saddle-point limit assumes that every configuration allowed by the strong-interaction conservation laws can be reached; the frozen limit is the opposite assumption, appropriate if the hadronic composition does not fluctuate and the droplet cannot exchange conserved charges with its surroundings by diffusion within the time it exists. A detailed discussion and a preliminary framework are provided in~\cite{Guerrini:2025ojt}.

\begin{figure*}
    \centering
    \includegraphics[width=\textwidth]{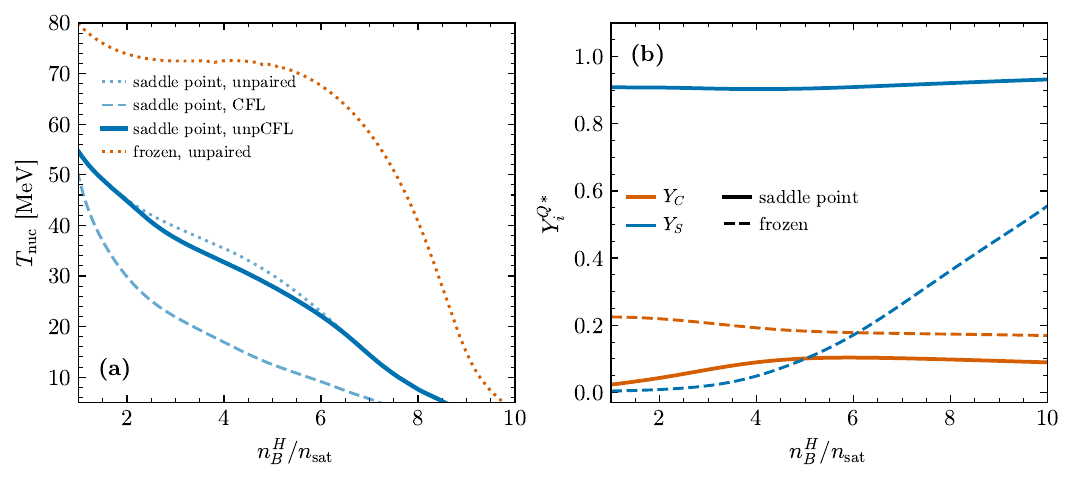}
    \caption{Frozen-flavor versus saddle-point composition, for Set~A under local charge neutrality and $\YLH=0.25$. (a) Nucleation temperature $\Tnuc$ versus $\nBH/\nsat$ at $\sigma=100\MeVfm$, for the frozen unpaired droplet (orange) and for the saddle-point droplet in the unpaired, CFL and unpCFL prescriptions (blue; dotted, dashed and solid respectively). 
    (b) Composition of the critical droplet, $Y_C^{\Qs}$ and $Y_S^{\Qs}$, versus $\nBH/\nsat$ at $T=30\MeV$, for the saddle-point (solid) and frozen (dashed) prescriptions.}
    \label{fig:app_frozen}
\end{figure*}

The difference between the two is not a modest quantitative shift, see Fig.~\ref{fig:app_frozen}.
Panel~(a) shows the nucleation temperature $\Tnuc(\nBH)$ at $\sigma=100\MeVfm$ for the frozen unpaired droplet and for the three saddle-point prescriptions. The frozen curve lies far above all of them, with a difference of several tens of $\MeV$ that closes only at much higher densities. 

Panel~(b) shows why, by displaying the composition of the critical droplet itself as a function of $\nBH$ at $T=30\MeV$. The frozen droplet, by construction, is pinned to the strangeness of the bulk average hadronic phase, which is negligible at low density and increases as the hyperon population builds up. It also carries a much larger $Y_C^{\Qs}$, which costs additional free energy. 

\end{document}